\documentclass[11pt]{article}
\usepackage{geometry}               
\usepackage{graphicx}
\usepackage{amssymb}
\usepackage{epstopdf}
\title{A New Constraint on Effective Field Theories  of the QCD Flux Tube}

\author{M. Baker\\  {\it Dept of Physics, University of Washington} \\  {\it P. O. Box 351650, Seattle WA 98195, USA}}
\date{}                                     

\begin{document}
\maketitle

\sf

\begin{abstract}
{ Effective magnetic $SU(N)$ gauge theory with classical  $Z_N$ flux tubes  of intrinsic width $\frac{1}{M}$ is an effective field theory of the long distance quark-antiquark interaction
in $SU(N)$ Yang-Mills theory. Long wavelength fluctuations of the $Z_N$ 
vortices  of this  theory lead to an effective string 
theory. In this paper we clarify the connection between effective field theory and effective string theory and we propose a new constraint on these vortices.

We first examine  the impact of string fluctuations on the classical dual superconductor description of confinement. At  inter-quark distances $R\sim \frac{1}{M}$
the classical action for a straight flux tube determines the
heavy quark potentials.   
At distances $R \gg  \frac{1}{M}$  fluctuations of the flux tube axis $\tilde{x}$ give rise to an 
effective string theory with an
action $S_{eff} (\tilde{x})$,   the
classical action for a curved flux tube, evaluated 
in the limit $\frac{1}{M} \rightarrow 0~$. This action is
equal to the Nambu-Goto action.

These conclusions are independent of the details of the   
$Z_N$ flux tube.  Further, we assume the QCD flux tube satisfies the additional constraint:
$$\int_0^\infty  r dr \frac{T_{\theta \theta} (r)}{r^2} =0,$$
where $\frac{T_{\theta \theta}(r)}{r^2}$ is the value of the  $\theta\theta$ 
component of the stress tensor  at a distance $r$ from the axis of an infinite  flux tube.
Under this constraint the string tension $\sigma$ equals the force on a quark in
the chromoelectric field $\vec{E}$  of an infinite straight flux tube, and
the Nambu-Goto action 
can be represented in terms 
of the chromodynamic fields  of effective magnetic $SU(N)$ gauge theory, 
yielding a field theory interpretation of  effective string theory.}
\end{abstract}

 \newpage

\sf

\section { Introduction}

\subsection {Background}

\subsubsection {Dual Superconductor Mechanism of Confinement}

In  the dual superconductor 
mechanism for confinement \cite{Nambu, Mandelstam, tHooft} 
a dual Meissner effect confines color electric flux to  a narrow 
flux tube connecting a quark-antiquark pair, and as a consequence, the 
energy of the pair increases linearly with their separation, confining the 
quarks in hadrons.

The Abelian Higgs model is an example
of a relativistic field theory having confining vortex solutions \cite{NO}.
The $U(1)$ gauge symmetry is completely broken by scalar Higgs fields $\phi$,
which vanish on the axis of  the flux tube and  increase to their non-vanishing vacuum value $\phi_0$ at large distances
from the vortex.
Interpreting the $U(1)$ symmetry as a magnetic  gauge symmetry coupling
"dual" potentials to magnetically charged Higgs fields with magnetic
coupling constant $g$, the flux tube then carries electric 
flux $\frac{2\pi}{g}$ confining a "quark" and an "antiquark" attached to
its ends \cite{Nambu}.

\subsubsection{Effective Field Theory of Dual Superconductivity}

Spontaneously broken magnetic $SU(N)$ gauge theory,
describing non-Abelian "dual" potentials $C_\mu$  coupled to magnetically
charged adjoint representation scalar Higgs fields $\phi_i $,
provides a non-Abelian example of an effective field theory of the long distance  
quark-antiquark interaction in $SU(N)$ Yang-Mills theory \cite{BBZ:1990,BBZ:1991}.   "Dual" potentials or "electric vector potentials" $C_\mu$
were first defined kinematically by Mandelstam \cite{Mandelstam2}
in terms of 't Hooft loops \cite{tHooft2}, operators which create 
vortices of magnetic flux. 
The spatial components of the field tensor $ G_{\mu \nu}$,
constructed from the potentials $C_\mu$, determine the color electric field
$\vec{E}$ and the space-time components the color
magnetic field $\vec{B}$.  The fields $\vec{E}$ and $\vec{B}$ 
evaluated at the position of the 
quarks can be identified with the corresponding chromodynamic
fields of the underlying $SU(N)$ Yang-Mills theory \cite{Nora}.    

This effective field theory possesses (i) the  $SU(N)$ symmetry of Yang-Mills theory
and (ii) the same low energy spectrum ;  i.e., it contains no massless 
particles and has $Z_N$ electric flux tube solutions.  The gauge coupling constant is denoted $g_m$, and the  magnitude of the vacuum value of the Higgs field is denoted $\phi_0$.
 The mass  $M \sim g_m \phi_0$ of the vector particle arising from the non-Abelian 
Higgs mechanism determines the flux tube 
intrinsic width  $\frac{1}{M}$.  The energy per unit length of the classical flux tube, the string tension $\sigma    \sim  \#~ \frac{M^2}{g_m^2}$. 

\subsubsection{ Effective String Theory from Effective Field Theory} 

When the distance $R$ between the quark and antiquark is much larger
than $\frac{1}{M}$, long wavelength
fluctuations of the $Z_N$ vortices become important and lead to an 
effective string theory of these fluctuations \cite{Baker+Steinke}.
The action $S_{eff} (\tilde{x})$ of this effective string theory
equals $S^{class} (\tilde{x})$, the classical action for a
curved vortex sheet $\tilde{x}$, evaluated in the
limit $\frac{1}{M} \rightarrow 0$. This action equals the Nambu-Goto 
action with the classical string tension.
$S_{eff} (\tilde{x})$ is then equal to the Nambu-Goto action.

\subsection {Effective String Theory}

The long distance $q \bar{q}$ interaction 
is usually described by effective string theory \cite{Luscher, Luscher 1, Luscher3} with an action $S_{eff } (\tilde{x})$ in which
the string tension $\sigma$ is an independent parameter.
The heavy quark potential  $V(R)$ is an expansion in powers 
of $\frac{1}{\sigma R^2}$. The leading terms in this expansion are  
the linear potential and the universal
L{\" u}scher term  \cite{Luscher}: 
\begin {equation}
V(R) = \sigma R - \frac{\pi}{12R} +  .\,.\,.
\label{first second}
\end{equation}

Effective string theory has since been developed extensively.
It has been shown \cite{ aharony1,  aharony2}
that consistency with Poincare symmetry requires that the  expansion 
of the ground state heavy quark potential in powers 
of $\frac{1}{\sigma R^2}$ coincides to order $\frac{1}{R^5}$ with the 
 potential generated by the Nambu-Goto action.  (Boundary terms in $S_{eff}$ give corrections of order $\frac{1}{R^4}$.)

Since the Nambu-Goto action is the action of the effective string
theory obtained from effective field theory, this result
implies that effective field
theory accounts for the contributions of string fluctuations
to the ground state heavy quark potential to order $\frac{1}{R^5}$.
Higher order terms in this long distance expansion are not taken into 
account by effective field theory.

\subsection {The Goal of this Paper}

The purpose of this paper is twofold: (i)
to clarify the connection between effective field theory and effective string theory, and  (ii) to propose a new constraint on the structure of the QCD flux tube.

\subsection {Impact of String Fluctuations on the Flux Tube Picture}

We first examine the impact of string fluctuations on the classical description of confinement. At distances $R\sim \frac{1}{M}$, 
the classical action for a straight flux tube determines the
heavy quark potential $V(R)$.
 Calculations \cite{BBZ:1995, antonio}  of heavy quark potentials in the model introduced in  \cite{BBZ:1990} were consistent with early lattice simulations \cite{bali} 
with  $M  \sim 2 \sqrt{\sigma}$ ~\cite{BBZ:1997}.\footnotemark 

\footnotetext[1] {Since $SU(3)$ lattice simulations \cite{lucini2002}  of pure gauge theory yield a  deconfinement temperature $T_C \approx 0.65 \sqrt{\sigma} \sim \frac{M}{3}$ there is an interval of temperatures  where we expect that effective magnetic gauge theory is also applicable in the deconfined phase \cite{baker2008}.}
 At distances $R \gg \frac{1}{M}$, 
where corrections due to string fluctuations become important, effective string theory determines the heavy quark potential. In an intermediate range of distances 
between approximately  $ \frac{1}{M}$ and $ \frac{2}{\sqrt{\sigma}}$ 
both the flux tube intrinsic width and the effect of string fluctuations
must be taken into account. Both effects were considered in the recent
analysis \cite{pedro} of lattice simulations  of  field distributions surrounding a quark-antiquark pair for a range of values of their separation.

\subsection { A  Constraint  on the Confining Flux Tubes}

\begin{figure}[htbp]
\begin{center} 
\includegraphics[width=3in]{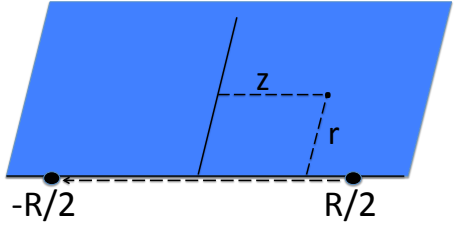}
\caption{ Half plane passing through the axis of the
flux tube. Eq. (\ref{balance}) is the condition that the torque 
per unit length acting across any such (r, ~z) plane must vanish as $R \rightarrow \infty$.}
\label{fluxtube}
\end{center}
\end{figure}

The motivation for  our constraint  is based on the following expression for the string tension $\sigma$,  derived in section (\ref{Infinity}) and  valid for any form of the Higgs potential $V (\phi_i)$ for which the $\frac{SU(N)}{Z_N}$ symmetry of the effective field theory is completely broken:

 \begin{equation}
 \sigma 
 = \,\,2\, tr \left [ -\frac {2 \pi}{g_m} Y\, \vec{E} (r=0) \right] \cdot \hat{e}_z   
 -\,2 \pi \tau,
 \label{sigmaWprime}
 \end{equation}

where

\begin {equation} 
2\, tr \left [ -\frac {2 \pi}{g_m} Y\, \vec{E} (r=0) \right]  \equiv \vec{F}
\label{force}
\end{equation}

\noindent is  the chromodynamic force on a quark in the color field $\vec{E} (r=0)$ on the  axis of an infinite $Z_N$ flux tube.
(Both the quark color charge  $-\frac{ 2 \pi}{g_m} Y$ and the color field $\vec{E} (r=0)$ have $N$ components and the trace in (\ref{force})  is a sum of the products of these components.)
$\tau$ is the torque per unit length on any $r, z$ half plane $(\theta =$ constant, $r > 0$) passing through the axis of the flux tube ( Fig. \ref{fluxtube})  and is given by

\begin {equation}
\tau \equiv  \int_0^\infty  r dr \frac{T_{\theta \theta} (r)}{r^2} ,
\label{tau}
\end{equation}

\noindent where $\frac{T_{\theta\theta} (r)}{r^2}$ is the value of the $\theta \theta$ component of the stress tensor at a distance $r$ from the flux tube axis (the z-axis). 
$T_{\theta \theta} (r)$ defines an azimuthal pressure $p(r)$,

\begin {equation}
p(r) \equiv \frac{T_{\theta\theta} (r)}{r^2} ,
\label{pressure}
\end{equation}

\noindent and $\tau$ is the radial moment of this pressure distribution.

Eq (\ref{sigmaWprime})  is the work-energy relation for a flux tube.  
The work  per unit length needed to move a quark along the flux tube axis is $\vec{F} \cdot \hat{e}_z$ . The work  per unit length required to remove the field energy in a sector $\Delta \theta$ of the flux tube while maintaining the quark-antiquark separation is   $-\Delta \theta \tau$, so that  $- 2 \pi \tau$ is  required to remove all the field energy. The flux tube energy per unit length $\sigma$ is then the sum (\ref{sigmaWprime}) of these two contributions to the work per unit length.

The torque per unit length $\tau$ is a new long distance parameter of effective field theory relating the  string tension to the color field on the flux tube axis via (\ref{sigmaWprime}).   We assume that the value  $\tau =0 $ characterizes the structure of the QCD flux tube, distinguishing it from the flux tubes arising from other field theories: i. e.,

\begin {equation}
\tau  \equiv  \int_0^\infty  r dr \frac{T_{\theta \theta} (r)}{r^2}  = 0. 
\label{balance}
\end{equation}

\noindent If the constraint  (\ref{balance})  is met, then by  (\ref{sigmaWprime}) the string tension $\sigma$ is  equal to the force on a quark in the 
chromoelectric field $\vec{E} (r=0)$ on the  axis of an infinite flux tube:

\begin {equation}
\sigma  = 2 tr \left [ -\frac{2 \pi}{g_m} Y \vec{E}(r=0) \right ]  \cdot \hat{e}_z  =  \vec{F} \cdot \hat{e}_z.
\label{newsigma}
\end{equation}

\noindent  Our conjecture is that the equivalent conditions (\ref{balance}) and (\ref{newsigma})  characterize the QCD flux tube.

Condition (\ref{balance}) means that
the work per unit length  required to remove the  field energy available after a quark-antiquark pair have been separated by a distance $R$ approaches $0$ in the limit $R \gg \frac{1}{M}$.   Then the long distance heavy quark potential  $\sigma R$ becomes equal to the work  $\vec{F} \cdot R \hat{e}_z$ needed to separate the quark-antiquark pair a distance $R$ in the  field $\vec{E} (r=0)$ on the axis of an infinite flux tube,  which is condition (\ref{newsigma}).

\subsection {Outline of This Paper}

In section \ref{Effective} we provide the background and notation
used in the paper, and
we discuss   $Z_N$ flux tubes
and their coupling to a quark-antiquark pair. 
We review the 
transition  from effective field theory to effective 
string theory \cite{Baker+Steinke} in section \ref{effectivefieldtheory}, and  discuss
the interplay between the width due to string fluctuations
and the intrinsic width of the flux tube.

In section \ref{Classical Action}  we derive a generalization of (\ref{sigmaWprime}) to curved vortex sheets $\tilde{x}$ to obtain an expression for $S^{class} (\tilde{x})$, the classical action for the
vortex sheet $\tilde{x}$ determining the action of the effective string theory. We use this expression in section \ref{Speculation}, where we impose our constraint (\ref{balance}) on flux tubes. Making use of Poincare invariance, we then obtain a representation of the Nambu-Goto action as an integral over  the chromodynamic force on the vortex sheet.  This representation is the generalization of (\ref{newsigma}) to  curved vortex sheets, and gives a field theory interpretation of effective string theory.

In section \ref{specific} 
we examine this  picture in a particular $SU(3)$ example \cite{BBZ:1990}
where
explicit classical $Z_3$ flux tube solutions  have been found .
The constraint  $\tau=0$
fixes the value of a parameter $\kappa$ in the Higgs potential 
of the non-Abelian theory.  This parameter plays the role 
of the Landau-Ginzburg parameter of the Abelian Higgs model. In the Summary  we discuss the possibility of testing the conjecture (\ref{balance}) using
lattice simulations. 

\section {Effective Magnetic $SU(N)$ Gauge Theory }
\label{Effective}

We consider effective field theories  coupling 
magnetic $SU(N)$ gauge potentials $C_\mu$ to adjoint representation 
scalar fields $\phi_i$.  The gauge coupling constant  is $g_m$ .
The magnetic gauge potentials $C_\mu$ and Higgs fields $\phi_i$ are 
elements of the Lie Algebra of SU(N).
We use a time-like metric: $C_{\mu} = (C_0,-\vec{C}) = \sum_a C_\mu^a T_a, \,\, \phi_i = \sum_a \phi_i^a T_a$, where the $T_a = \lambda_a/2$ are the fundamental representation generators normalized so that
\begin {equation}
 2 tr T_a T_b = \delta _{a\,b}.
\label{repgen}
\end{equation}

The effective Lagrangian is
\begin{equation}
{\cal{L}}_{eff} (C_\mu, \phi_i)=  2 tr \left( -\frac{1}{4} G^{\mu \nu} G_{\mu \nu} + \frac{1}{2} ({\cal{D}}_\mu \phi_i)^2 \right ) - V(\phi_i),
\label{Leff}
\end{equation}
with 
\begin{equation}
G_{\mu \nu} = \partial_\mu C_\nu - \partial_\nu C_\mu - i g_m [C_\mu, C_\nu],
\label{Gmunu}
\end{equation}
and
\begin {equation}
{\cal{D}}_\mu \phi _i= \partial_\mu \phi_i - i g_m [C_\mu, \phi_i].
\label{Dmu}
\end{equation}
The components of the field tensor $G^{\mu\nu}$ define color 
electric and magnetic fields $\vec{E}$ and $\vec{B}$: 
\begin {equation}
E^k = \frac{1}{2} \epsilon_{k l m}G^{l m} ,\, \, B^k = G^{k0}  . 
\label {EB}
\end{equation}

\noindent $V (\phi_i)$ is an $SU(N)$ invariant Higgs potential which has 
an absolute minimum at a non-vanishing
value $\phi _{i \, 0}$ of the Higgs fields such that in the 
confining vacuum,
\begin{equation}
C_{\mu} = 0, \,\,\, \phi_i =\phi_{i \, 0} \,\, ,
\label{phi}
\end{equation}
the $\frac{SU(N)}{Z_N}$ symmetry is completely broken and all
particles become massive.  
The number of  Higgs fields and the form of the Higgs potential are otherwise unspecified.
In section \ref{specific} 
we will write down a specific $SU(N)$ Higgs potential for which 
explicit $Z_3$ flux tube solutions
were found.\footnotemark

\footnotetext[2]{  For a general discussion of magnetic vortices
in non-Abelian gauge theory see \cite{konishi}.}

\subsection { $Z_N$ Electric Flux Tubes } 
\label{Electric}
Effective magnetic gauge theory has electric $Z_N$ flux tube 
solutions for which,
 at large distances $r$ from the flux tube axis,
 $C_{\mu} $ and $\phi_{i} $  approach
a gauge transformation $\Omega (\theta)$ of the
vacuum fields \cite{konishi} :

\begin {equation}
C_\mu \rightarrow \frac{i}{g_m} \Omega^{-1} (\theta) \partial_\mu \Omega(\theta) ,\,\,\,
\phi_i(x) \rightarrow  \Omega^{-1} (\theta) \phi_{i0} \Omega(\theta).
\label{vacsoln}
\end{equation}

\noindent In order that the Higgs field be single valued on any path encircling
 the z-axis the matrix $\Omega^{-1}(\theta = 2 \pi)  \Omega (\theta = 0) $
must commute with  all the $\phi_{i0}$ 
and, since the gauge symmetry is completely broken, must be an element of $Z_N$: 
 $ \Omega(\theta = 2 \pi) = exp (2\pi i k/N) \, \Omega(\theta = 0),\,\, k = 0,1,2,....N-1$. 

We can choose a gauge where $  \Omega $ is Abelian.  For a
$Z_N$ flux tube with $k=1$ we take

\begin{equation}
 \Omega(\theta) = exp( i \theta Y) \, ,
\label {Utheta}
\end{equation}
where $Y$ is a diagonal matrix.  Its first $N-1$ elements = $1/N$ and
its Nth element $= - (N-1)/N$. ( There are $N$ physically equivalentCoupling
choices for $Y$ related to each other by a gauge transformation 
\cite{konishi}). With the choice (\ref{Utheta}) for $\Omega (\theta)$,

\begin{equation}
C_\mu \rightarrow \frac{-\partial_\mu \theta}{g_m}Y, \,\,
 as \,\, r \rightarrow \infty, \label{C}
\end{equation}
so that
\begin {equation}
\vec{C} \rightarrow  \frac{1}{g_m r} \hat{e}_\theta  Y, \,\,as\,\, r \rightarrow \infty.
\label{vecC}
\end{equation}

\noindent Integrating $\vec{C}$ around a path at large r surrounding the z-axis gives

\begin {equation}
exp (i g_m \oint \vec{C} \cdot d \vec{l}) =  exp(2 \pi  i Y )= exp( \frac{2 \pi i}{N}),
\label{oint}
\end{equation}
reflecting the one unit of $Z_N$ electric flux passing through the $x y$ plane. 

We assume that there is a classical solution where the
gauge potential $\vec{C}$ is everywhere proportional to the matrix $Y$: 
\begin{equation}
\vec{C} = C(r) {\hat{e}}_\theta Y .
\label{Cft}
\end{equation}
The flux tube electric field (\ref{EB}) also lies along
the $Y$ direction in color space:
\begin{equation} 
\vec{E} (r) = -\vec{\nabla}\times\vec{C}  \,\,=
-\frac{1}{r}\frac{d(rC(r))}{dr} Y  {\hat{e}}_z ,
\label{vecE}
\end{equation}
The Higgs fields $\phi_i$ have
the form : 
\begin{equation}
\phi_i =  \Omega^{-1}  (\theta ) \phi_i (r) \Omega (\theta), ~~~ \rm{where} ~~~\phi_i (r) \rightarrow \phi_{i0},~~ r \rightarrow \infty.
\label {phiofomega}
\end{equation}

In order that the flux tube have finite energy the Higgs fields $\phi_i$  for which $[C_\mu, \phi_i] \ne 0$
must vanish on the flux tube axis $r=0$. 

The vector mass $M$ generated by the Higgs condensate, which  determines the intrinsic width  $\frac{1}{M}$ of the flux tube, is obtained by replacing $\phi_i$  by  $\phi_{i 0}$ and $C_\mu$ by $Y$ in
 (\ref{Leff}),  is
\begin {equation}
M^2 = g_m^2 \sum_i\frac{ 2 tr[ iY, \phi_{i 0} ]^2}{2 tr Y^2}.
\label{explicitM2}
\end{equation}

\subsection {Coupling of $Z_N$ Flux Tubes to Quarks}
\label{Coupling}

Classical $Z_N$ vortices of magnetic $\frac{SU(N)}{Z_N}$ gauge 
theory  carrying  one unit of $Z_N$ flux  
couple to a quark-antiquark pair in the fundamental representation 
of  $SU(N)$  via a {\cal{D}}irac string $G_{\mu \nu}^s$,  carrying color charge $\frac{2\pi}{g_m} Y$, which is non vanishing on some line
 connecting the pair. 
 
  Long wave length fluctuations of the axis  of the flux tube sweep
out a space-time surface $\tilde{x}^\mu (\sigma, \tau)$  bounded by the loop $\Gamma$ formed by the world lines of the quark and antiquark at the ends of the vortex.  
We assume that the classical solution $C_\mu$ having a vortex on the sheet $\tilde{x}^\mu (\sigma, \tau)$ 
is also proportional to the matrix $Y$: 
\begin {equation}
C_\mu = C_\mu (x, \tilde{x})\,Y.
\label{Cmu}
\end{equation}
(For $SU(3)$ we have obtained an 
explicit solution (\ref{newphi}), (\ref{Cmu2}) where $C_\mu$ has the form 
(\ref{Cmu}) with $Y = \frac{\lambda_8}{\sqrt3}$.)

The Higgs fields  
$\phi_i$  for which $[Y, \phi_i] \ne 0$
contribute to the magnetic current density, the source of the potential 
$C_\mu$, and
must vanish on $\tilde{x}^\mu (\sigma, \tau)$. 
We choose a gauge where the surface swept out by the Dirac string coincides with the vortex sheet $\tilde{x}^\mu (\sigma, \tau)$.  The corresponding Dirac polarization tensor $G^s_{\mu \nu} = G^s_{\mu \nu} (x, \tilde{x})$ is  \cite{Dirac}
\begin {equation}
G^s_{\mu \nu} (x, \tilde{x}) = -\frac{1}{2} \epsilon_{\mu \nu \alpha \beta} \int d \tau \int d \sigma \sqrt {-g}\,t^{\alpha \beta} \delta (x - \tilde{x} (\sigma, \tau)) \frac{2\pi}{g_m}Y,
\label{Gmunuagain}
\end{equation}
where $g$ is the determinant of the induced metric $g_{ab}$,
\begin {equation}
g_{ab} = \frac{\partial \tilde{x}^\mu}{\partial \xi^a} \frac {\partial \tilde{x}_\mu}{\partial \xi^b},~~~~~\xi^1 = \tau,\,\,  \xi^2 = \sigma,
\label{gab}
\end{equation}
$\tilde{x}^\mu (\xi) \equiv \tilde{x}^\mu (\sigma, \tau)$ 
is a parameterization of the vortex sheet 
and
\begin {equation}
t^{\alpha \beta} = \frac{1}{\sqrt{-g}} ( \frac{\partial \tilde{x}^\alpha}{\partial \tau} \, \frac{\partial \tilde{x}^\beta}{\partial \sigma} \,-\, \frac{\partial \tilde{x}^\alpha}{\partial \sigma}  \frac{\partial \tilde{x}^\beta}{\partial \tau} )
\label{tab}
\end{equation}
is the tensor specifying the orientation of the surface $\tilde{x}^\mu (\sigma, \tau)$ in four dimensional space time.
It is invariant under a reparameterization of the surface $\tilde{x}^\mu$ and normalized so that $t^{\alpha \beta} t_{\alpha \beta} = - 2.$

The action $S[ C_\mu, \phi_i]$  describing field configurations having a vortex on the sheet $\tilde{x}^\mu (\sigma, \tau)$ coupling the dual potential (\ref{Cmu}) to $G^s_{\mu \nu}$ is

\begin {equation}
S[ C_\mu, \phi_i] =  \int dx\, {\cal{L}}_{eff}( C_\mu, \phi_i,~G^s_{\mu \nu} (x, \tilde{x})),  
\label{action1}
\end{equation}
where the Lagrangian ${\cal{L}}_{eff}(C_\mu, \phi_i, G^s_{\mu \nu} (x, \tilde{x}))$ 
is  obtained by replacing, in the Lagrangian 
(\ref{Leff}), 
the dual field strength tensor $G_{\mu \nu}$ by
\begin {equation}
G_{\mu \nu} = \partial_\mu C_\nu - \partial_\nu C_\mu + 
G_{\mu \nu}^s (x, \tilde{x}).
\label{newGmunu}
\end{equation}
(The non-linear term $-i g_m [C_\mu,~C_\nu]$ in (\ref{Gmunu}) does not contribute to the field tensor (\ref{newGmunu}) in the gauge where the classical solution (\ref{Cmu}) is Abelian. )

After having partially fixed the gauge by the choice (\ref{Cmu}) the action (\ref{action1}) has a residual 
invariance under the  Abelian gauge transformation 
$\Omega = exp[i \Lambda (x) Y]$ :
\begin {equation}
C_\mu \rightarrow C_\mu  - \frac{1}{g_m} \partial_\mu \Lambda (x) Y ,\,\,\,
\phi_i \rightarrow  \Omega^{-1}  \phi_i \Omega. 
\label{abgt}
\end{equation}

\section {Effective Field Theory}
\label{effectivefieldtheory}
The 't Hooft loop acting in effective magnetic gauge theory creates a vortex of electric flux, and its expectation value determines the Wilson loop  $W (\Gamma)$ of Yang-Mills theory, calculated in magnetic gauge theory. $W (\Gamma)$ is the partition function of  the effective field theory in the presence of a Dirac string; i. e.,
$W (\Gamma)$ 
is a path integral over all field configurations $C_\mu, \phi_i$
having a vortex on any surface $\tilde{x}^\mu (\sigma, \tau)$ bounded by
the loop $\Gamma$ formed from the world lines of a quark--antiquark pair \cite{Nora}:

\begin {equation}
W(\Gamma) = \int{\cal{D}} C_\mu{\cal{D}} \phi_i \,exp(i[ S(C_\mu, \phi_i) + S_{g f}]),
\label {WGamma}
\end{equation}
where $S(C_\mu, \phi_i)$ is the action (\ref{action1}), while
$S_{gf}$ is a gauge fixing term. The path integral (\ref{WGamma}) is cut off at a scale $\Lambda$, which 
must be less than 
the mass of the  lightest glue ball, the lightest particle which has been integrated out in 
obtaining ${\cal L}_{eff}$.
$\Lambda$ must also  be somewhat greater than the vector mass $M$ in order to resolve
distances of the order of the flux tube radius.

Identification of the partition function (\ref{WGamma}) and the Wilson loop $W (\Gamma)$ implies that the expectation value of the field tensor $G_{\mu \nu}$ at the position of the quarks can be identified with the corresponding expectation values  of the color fields of Yang Mills theory \cite{Nora}. (For a static quark-antiquark pair separated by a distance $R$,
the loop $\Gamma$  is a rectangle in the $z~t$ plane and $W (\Gamma)$, evaluated in the limit as the elapsed time $T \rightarrow \infty$, determines the static heavy quark potential $V(R)$.)

We now briefly summarize the results of  \cite {Baker+Steinke},
where the field theory path integral (\ref{WGamma})
was transformed into a partition function of an effective string theory 
of vortices.

\subsection{ From Effective Field Theory  to Effective String Theory }
\label{Effective String 1}

To transform  $W(\Gamma)$ into 
a path integral over vortex sheets $\tilde {x}^\mu (\xi)$ 
we carry out the functional integration in two stages:

\begin {enumerate}
\item 
We first fix the location $\tilde{x}^\mu (\xi)$ of a particular vortex.

We integrate over field configurations in  (\ref{WGamma}) having a vortex 
on this particular surface.  
The integration over these configurations is proportional to $e^{i S_{eff} ( \tilde{x})}$, defining  
the action of the effective string theory $S_{eff} (\tilde{x})$, and the constraint on this integration introduces a Fadeev-Popov determinant into the functional integral (\ref{WGamma}).

The one loop calculation of (\ref{WGamma}) in an expansion around the classical solution includes a contribution from field modes generated by moving the position of the vortex. This contribution is cancelled by the  Fadeev-Popov determinant, so that only massive modes contribute to the one loop integration. Since
(\ref{WGamma}) is cut off at a scale $\Lambda$ which is only slightly larger than
the mass $M$ of the vector particle, the lightest particle
in the effective field theory,  the one loop corrections to $W(\Gamma)$ are negligible at the distance scales  $\sim \frac{1}{M}$ described by effective field theory.
$ S_{eff} (\tilde{x})$ can  then be approximated by $S^{class} (\tilde{x})$,
the value of the action at the classical configuration 
$( C^{class}_\mu  (x, \tilde{x}),\,\phi^{class}_i (x, \tilde{x} ) )$ 
minimizing the action (\ref{action1}) for a fixed position 
$\tilde{x}^\mu (\xi)$ of the vortex:

\begin {equation}
S_{eff} (\tilde{x}) \approx S (\tilde{x}, C^{class}_\mu (x, \tilde{x}), \phi^{class}_i (x, \tilde{x} ) ) \equiv S^{class}(\tilde{x}).
\label{newSeff}
\end{equation}


\item
We then integrate over all surfaces $\tilde{x}^\mu (\xi)$. 

We choose a particular 
parameterization of
$\tilde{x}^\mu $ in terms of the amplitudes
$f^1 (\xi)$ and $f^2 (\xi)$ of the two transverse fluctuations of the
vortex sheet,
\begin {equation}
\tilde{x}^\mu = x^\mu ( \xi,~f^1 (\xi),~ f^2 (\xi)),
\label {tildex}
\end{equation}
This gives $W(\Gamma)$ the form of a path integral of an effective string theory
of vortices:

\begin {equation}
W(\Gamma) = \int  \textit{D} f^1 \textit{D} f^2 \Delta\, exp(i S_{eff} (\tilde{x}),
\label{Wagain}
\end{equation}
where
\begin {equation}
\Delta \equiv {\cal{D}}et \left [ \frac{\epsilon_{\mu \nu \alpha \beta}}{\sqrt{-g} } \frac{\partial x^\mu}{\partial f^1}  \frac{\partial x^\nu}{\partial f^2} \frac{\partial \tilde{x}^\alpha}{\partial \xi^1} \frac{\partial \tilde{x}^\beta}{\partial \xi^2 }\right ]
\label{deltafp}
\end{equation}
is the determinant produced by gauge fixing the reparameterization 
symmetry.  The path integral representation (\ref{Wagain}) for $W(\Gamma)$ 
is invariant under  reparameterizations of the vortex sheet
$\tilde{x}^\mu(\xi)$, and is restricted to wavelengths longer 
than $\frac{1}{\Lambda}$.

\end{enumerate}

The action of the effective string theory $S_{eff} (\tilde{x})$ 
is the action (\ref{newSeff}) of the effective magnetic
gauge theory evaluated at a classical solution for a curved vortex  sheet
$ \tilde{x}$.
Since the contribution of string fluctuations to the heavy quark 
interaction determined by  the path integral 
(\ref{Wagain}) is applicable only for quark-antiquark 
separations $R\gg\frac{1}{M}$,
in this integral the action $S^{class} (\tilde{x})$ must be evaluated
in the limit  $\frac{1}{M} \rightarrow 0$.  I. e., $S_{eff} (\tilde{x}) = S^{class} (\tilde{x})|_{\frac{1}{M}=0}$.
In this limit $S^{class} (\tilde{x})$ depends only upon a single 
dimensional parameter, the classical string tension $\sigma$, and by
Poincare symmetry it must equal
the Nambu-Goto action $S_{NG} (\tilde{x}^\mu) $:

\begin {equation}
S^{class} (\tilde{x})|_{\frac{1}{M}=0}
=S_{NG} (\tilde{x}^\mu) \equiv -\sigma \int d^2 \xi  \sqrt{-g(\tilde{x}^\mu (\xi))}. 
\label {newaction}
\end{equation}

The action of the effective string theory obtained from effective field 
theory is then the Nambu-Goto action.  Since deviations 
from the Nambu-Goto action give contributions to 
the ground state heavy quark potential that fall off faster
than $\frac{1}{R^5}$ \cite {aharony1, aharony2}, effective field theory
describes the expansion of ground state heavy quark potential to order
$\frac{1}{R^5}$.  Higher order terms in this long distance expansion are
not taken into account by effective field theory and are not considered in
this paper.

With the use of analytic regularization  to
renormalize $S_{eff}(\tilde{x})$
no additional dimensional parameters appear in the resulting
static potential $ V(R)$, and the string tension $\sigma$ retains
its classical value as the energy per unit length of the
flux tube \cite{Dietz}.

For  a loop $\Gamma$ describing the motion of a quark-antiquark pair separated by a fixed distance and rotating with constant angular velocity, $W (\Gamma)$  determines the leading semi-classical correction to the classical formula for meson Regge trajectories \cite{Baker+Steinke2}.

\subsection {Width from String Fluctuations}
\label {fluctuation}

For distances much larger than  $\frac{1}{\sqrt{\sigma}}$, string fluctuations determine the flux tube width and
lead to a logarithmic increase of the mean square width $w^2 (R/2)$ of the flux tube at its midpoint  \cite{Munster}; 
\begin{equation}
w^2 (R/2) = \frac {d-2}{2 \pi \sigma}\, log \,\frac{R}{r_0}.
\label{logincrease}
\end{equation}
($\frac{1}{r_0}$ can be interpreted as the cutoff $\Lambda$ of
the effective field theory.  Fluctuations of wave lengths less than 
$\frac{1}{\Lambda}$ 
produce a divergent contribution to $w^2 (R/2)$.)

This prediction has been tested  by  very accurate lattice
simulations \cite{Gliozzi}  of the mean square
flux tube width in $d=2 + 1 $  SU(2) Yang-Mills theory
extending to distances $ R \approx \frac{36}{\sqrt{\sigma}}$.
These simulations gave excellent agreement with the prediction
(\ref{logincrease}) for distances $R > \frac{1.5}{\sqrt{\sigma}}$ with the
choice $r_0 =  \frac{0.364}{\sqrt{\sigma}}$ corresponding to a value of
$ \Lambda ~\sim~ 2.75 \sqrt{\sigma} \approx 1.4 M $.
However for distances $R < \frac{1.5}{\sqrt{\sigma }} $
the lattice simulations of $w^2(R/2)$  lie above the leading order 
prediction  (\ref{logincrease}) of effective string theory.
This excess may be interpreted as a manifestation 
of the flux tube intrinsic width at $q \bar{q}$ separations 
$R < \frac{1.5}{\sqrt{\sigma }} $.

\subsection {The Intrinsic Width of the Flux Tube}
\label{intrinsicwidth}

The intrinsic width produces 
an uncertainty 
of  order $\frac{1}{M}$ in the position of the vortex, so that for quark-antiquark separations $R \sim  \frac{1}{M}$
string fluctuations do not contribute to the path integral
(\ref{WGamma}).
The Wilson loop (\ref{WGamma}) can then be replaced by its 
minimum value, fixed by the value of the classical action for a flat
vortex sheet connecting a static quark-antiquark pair. $W (\Gamma)$ then yields  $V^{class} (R)$,
the heavy quark potential  
in the classical approximation.

\begin{figure}[htbp]
\begin{center}
\includegraphics[width=3in]{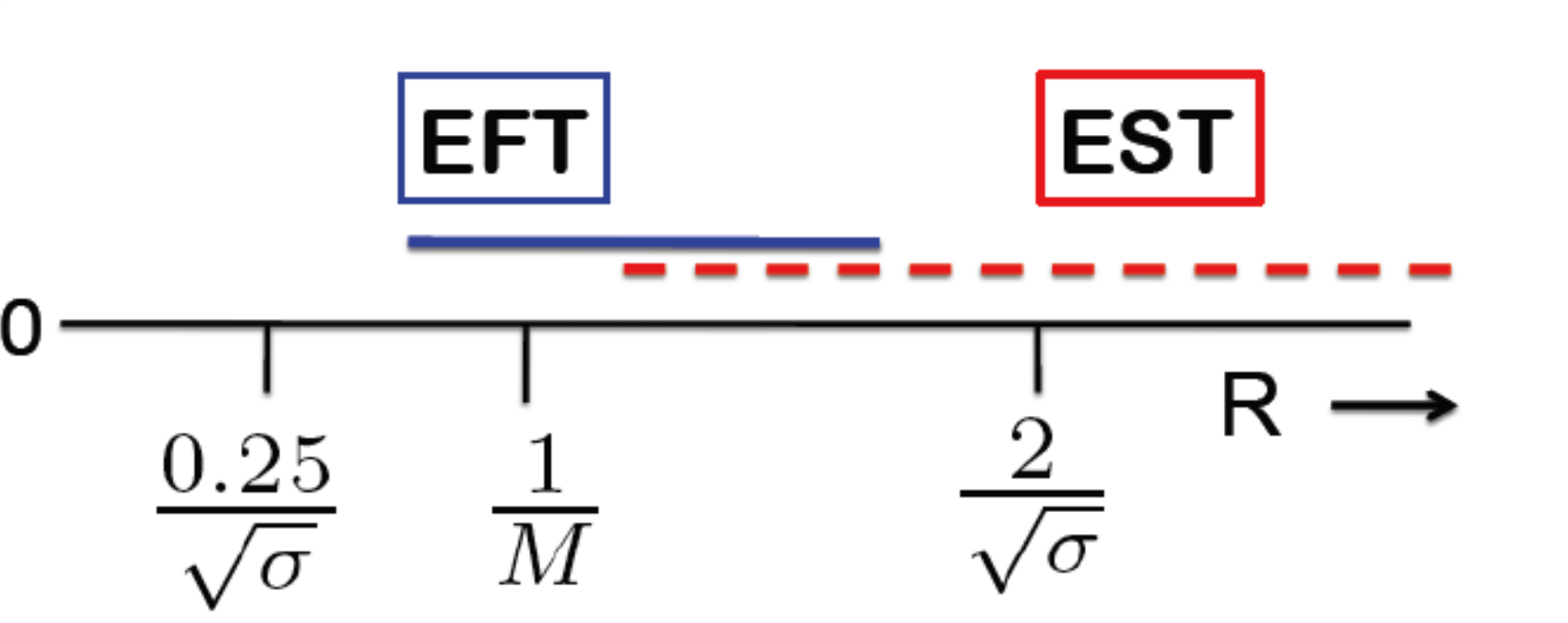}
\caption{Schematic showing  approximate domains of applicability of effective field theory (EFT) (solid blue line) and effective string theory (EST) (red dashed line).
}
\label{cartoon}
\end{center}
\end{figure}

Recent very accurate lattice simulations  \cite{pedro, simulations} of field and energy distributions in $SU(3)$ flux tubes find values of the intrinsic width characterizing these distributions that  corresponds to a  mass $M$ of approximately 900 MeV. 
Since  $ M \sim 2 \sqrt{\sigma}$, 
there is an interval of intermediate distances $R \sim \frac{1}{M}$ 
lying in the range 
where the predictions of effective field theory at the classical level 
are not washed out by string fluctuations.
(The lattice simulations of heavy quark potentials \cite{bali}, were carried out
at these distances.)
In this interval, denoted [EFT] in Fig. \ref{cartoon},
the classical flux tube picture should be manifest, 
while effective string theory should be used in  the distance range  $R > \frac{2}{ \sqrt \sigma}$ ( denoted [EST] in the Figure).

 Effective string theory must be used to fit more recent simulations  \cite{koma2} of heavy quark potentials for values of $R$ extending to $1.2 ~fm ~~ > \frac{2}{\sqrt{\sigma}}$.
In the intermediate  range of distances
depicted in Fig. \ref{cartoon} 
both the flux tube intrinsic width and the effect of string fluctuations
must be taken into account.

\section {The Classical Action for SU(N) Vortices}
\label{Classical Action}

We  now obtain a representation 
for the classical action of curved vortex sheets 
and a corresponding representation for flat sheets.
We will use these representations, 
together with Poincare invariance, to obtain information about the
classical action of a
general vortex sheet from the action  of a flat sheet.

Eq. (\ref{newSeff}) gives the action $S_{eff} (\tilde{x}^\mu)$ of the 
effective string theory as the action (\ref{action1})
of the effective field theory, evaluated at a classical solution
having a vortex at $\tilde{x}^\mu$.
To find the non-perturbative contribution to this action
we separate  $C_\mu$ into 
a perturbative contribution $C^D_\mu $  and a 
nonperturbative contribution $c_\mu $:
\begin {equation}
C_\mu = C^D_\mu + c_\mu = (C^D_\mu (x, \tilde{x}) + c_\mu (x, \tilde{x})) Y .
\label{effectivemag}
\end{equation}
The perturbative vector potential $C^D_\mu$ gives
the Maxwell field $G_{MAX}^{\mu \nu}$ of the external $q\, \bar{q}$ pair generated by the coupling of the dual potentials to $G^s_{\mu \nu} $ \cite{Dirac}: 

\begin {equation}
G_{MAX}^{\mu \nu}  = \partial^\mu C^{D\,\nu} - \partial^\nu C^{D\,\mu} + G^{s\,\mu \nu}.
\label{GMAX}
\end{equation}

The corresponding dual field tensor $G^{\mu\,\nu}$  assumes the form
\begin {equation}
G^{\mu \nu} =  G_{MAX}^{\mu \nu}  + G_{class}^{\mu \nu},
\label{Greplaced}
\end{equation}
where 
\begin {equation}
G_{class}^{\mu \nu} (x,\tilde{x}) = \partial^\mu c^\nu - \partial^\nu c^\mu = ( \partial^\mu c^\nu (x, \tilde{x})  - \partial^\nu c^\mu  (x, \tilde{x})  )Y,
\label{Gclass}
\end{equation}
is the non-perturbative field tensor  satisfying the classical equation of motion:
\begin {equation}
\partial_\nu G^{\nu \mu}_{class}  =   i g_m [\phi_i,{\cal{D}}^\mu \phi_i] \equiv  j^\mu, 
\label{cmueqofmotion}
\end{equation}
defining the magnetic current density $j^\mu$.
For consistency, the non-Abelian Higgs fields $\phi_i$ must have
a color structure such that  $j^\mu$ is also proportional to the matrix $Y$.

The action (\ref{action1}), evaluated at the classical solution,
separates into the sum:
\begin {equation}
S[ C_\mu, \phi_i, \tilde{x}^\mu] =  
S^{MAX}(\Gamma)+ S^{class} (\tilde{x}^\mu),
\label{sumS} 
\end{equation} 
where $ S^{MAX}(\Gamma)$ is (up to a color factor) the Maxwell action,
and  $S^{class}(\tilde{x}^\mu)$ is the non-perturbative 
contribution to the action:

\begin {equation} 
S^{class}(\tilde{x}) = \int dx \{ 2 tr \left [ - \frac{1}{4} G^{\mu \nu}_{class} G_{\mu \nu \,\,class} +  \frac{1}{2}({\cal{D}}_\mu \phi_i)^2 \right]  - V(\phi_i) \}.
\label{delta}
\end{equation}
(The classical action is related to 
the Hamiltonian:  $-\partial S/\partial t = H$ \cite{LL}.)

Using the equation of motion (\ref{cmueqofmotion}) in (\ref{delta}) to
write $2S^{class}(\tilde{x})$, and subtracting $S^{class}(\tilde{x})$
in the original form, gives
\begin {equation} 
S^{class}(\tilde{x}) = \int dx \{ 2 tr \left [ -\frac{1}{2}  
G_{\mu \nu}^{s} G_{class}^{\mu \nu} +   C_\mu j^\mu + \frac{1}{4} G^{\mu \nu}_{class} G_{\mu \nu \,\,class} +  \frac{1}{2}({\cal{D}}_\mu \phi_i)^2 \right]  - V(\phi_i) \}.
\label{deltaprime}
\end{equation}
(There is also a term on the right hand side of (\ref{deltaprime})
proportional to $G_{\mu\nu}^{MAX} G^{\mu \nu}_{class}$, which vanishes
after integration by parts and use of Maxwell's equations.)
Then, use of the identity
\begin{equation}
tr  \left (C^\mu j_\mu + \frac{1}{2} \left ({\cal{D}}_\mu \phi_i \right )^2 \right )
\equiv  tr \left (\frac{1}{2} (\partial_\mu \phi_i)^2 - \frac{[i g_m C_\mu , \phi_i]^2}{2} \right )
\label{identity}
\end{equation}
to rewrite  (\ref{deltaprime}) gives the following representation
of the classical action:

\begin {equation}
-S^{class}(\tilde{x}) =     \int dx 2 tr \left (\frac{1}{2} G_{\mu \nu}^s G_{class}^{\mu \nu}  \right ) + S_g(\tilde{x}) - S_\phi(\tilde{x}),
\label{newSclass}
\end{equation}
where
\begin {equation}
S_g (\tilde{x}) = \int dx 2 tr \left [ -\frac{1}{4} G^{\mu \nu}_{class} G_{\mu \nu \,\,class} - \frac{g^2_m [C_\mu, \phi_i]^2}{2} \right] ,
\label{Sg2}
\end{equation}
and
\begin {equation}
S_\phi (\tilde{x}) = \int dx [2 tr  (\frac{\partial_\mu \phi_i)^2}{2}  - V(\phi_i)  ].
\label{Sphi}
\end{equation}
The first term in (\ref{newSclass}),

\begin {eqnarray}
\label{Sgagain} 
 \int dx 2 tr  [\frac{1}{2} G_{\mu \nu}^s G_{class}^{\mu \nu}  ]& =&\\ \nonumber
  -\frac{1}{4} \int d \tau \int d \sigma \sqrt{-g}  \epsilon_{\mu \nu \lambda \alpha}\, 
 2 tr \left ( \frac{2 \pi}{g_m} Y 
  G^{\mu \nu}_{class} (x, \, \tilde{x} |_{x^\mu  
 = \tilde{x}^\mu (\sigma, \tau) }) \right)
 t^{\lambda \alpha} (\sigma, \tau)  
& \equiv &\int d \tau W (\tau),
\end{eqnarray}
the integrated  work 
required to separate the quark-antiquark pair along the 
vortex sheet $\tilde{x}^\mu (\sigma, \tau)$ in the fully 
developed  field $G^{\mu \nu}_{class} (x, \tilde{x})$,
and the second term,
\begin {eqnarray}
\label{Wfield}
S_g (\tilde{x}) - S_\phi (\tilde{x}) =\int dx \left(  2 tr [ -\frac{1}{4} G^{\mu \nu}_{class} G_{\mu \nu \,\,class} - \frac{g^2_m [C_\mu, \phi_i]^2}{2}]
-[2 tr \frac{(\partial_\mu \phi_i)^2}{2} - V(\phi_i)]  \right ), 
\end{eqnarray}
is the net additional integrated field energy available
from the process of creating the vortex sheet, i. e. it is
the difference between $-S^{class} (\tilde{x})$, the
integrated work needed to separate the quarks in the developing color  
fields, and the corresponding integrated work $\int d \tau W (\tau)$
in the fully developed  chromodynamic field $G^{\mu \nu}_{class}$. 

With a parameterization where 
$ \frac{\partial \tilde{x}^0}{\partial \sigma} |_\tau =0, $ (\ref{Sgagain})
takes the form:
\begin{equation}
 \int dx 2 tr [\frac{1}{2} G_{\mu \nu}^s G_{class}^{\mu \nu} ]
 = \int  d \tau \int d \sigma \frac {  \partial \vec {\tilde{x}}}{\partial \sigma} \cdot  \vec {F}_{class}  (\sigma, \tau, \tilde{x}) ( \frac{\partial \tilde{x}^0}{\partial \tau}),
\label{Wtau}
\end{equation}

where 
\begin{equation}
\vec {F} _{class}(\sigma, \tau, \tilde{x}) \equiv 2 tr \left [-\frac{2 \pi}{g_m} Y\,(\vec {E}_{class} (x,\,\tilde{x}) + \vec {v} \times \vec {B}_{class} (x,\,\tilde{x}) ) \right ]|_{x^\mu = \tilde{x}^\mu (\sigma, \tau)},
\label{F}
\end{equation}
\begin {equation}
E^k_{class} (x, \tilde{x})= \frac{1}{2} \epsilon_{k l m}\, ( G^{l m}_{class} (x,\,\tilde{x}) )\ ,\, B^k_{class}(x, \tilde{x}) =G_{class}^{k0} (x,\,\tilde{x}) 
\label {EBclass}
\end{equation}
are the classical chromoelectric and chromomagnetic fields, and
\begin{equation}
\vec{v} (\sigma, \tau) = \frac{\partial  \vec{\tilde{x}} (\sigma, \tau)}{\partial \tau}/ \frac{\partial \tilde{x}^0}{\partial \tau}
\label{vvec}
\end{equation}
is the velocity of the sheet.

\subsection {The Heavy Quark Potential in the Classical Approximation} 
\label{Relations}

The classical action $S^{class} (\tilde{x})$, evaluated for a flat vortex sheet connecting a static
quark at ${\vec{x}}_1=\frac{R}{2}{\hat{e}}_z$ and an antiquark at 
${\vec{x}}_2=-\frac{R}{2}{\hat{e}}_z$, determines  $V^{class} (R)$, the approximation to the heavy quark potential,  where string fluctuations are neglected.  For this sheet the components 
(\ref{Gmunuagain}) of $G_{\mu \nu}^s (x, \tilde{x})$ are given by
\begin{equation}
G^s_{k 0} =0,~ G^{s}_{ l m} = \frac{1}{2} \epsilon_{lmn} E^{sm}, ~{\vec{E}}^{s}  = -\frac{2 \pi}{g_m} \delta (x) \delta (y)[\theta(z+R/2)-\theta(z-R/2)] {\hat{e}}_z \,\,Y.
\label{Es}
\end{equation}
The vector potential (\ref{Cmu}) becomes
\begin{equation}
\vec{C}  = C (r , z ) \hat{e}_\theta\,Y, \,\,\,C_0 =0 .
\label{vecCagain}
\end{equation}
\noindent  The spatial components  of the tensor $G_{\mu \nu} $ (\ref{newGmunu}) 
yield the static chromoelectric field $\vec{E}$:
\begin{equation}
\vec{E} = -\vec{\nabla}\times\vec{C}  \,\,+  {\vec{E}}^{s} .
\label{Eagain}
\end{equation}
In cylindrical coordinates, $g_{00} =1,\,\,g_{zz} = g_{rr} = -1,
\,\,g_{\theta \theta} = - r^2$,   $g \equiv  det\, g_{\mu \nu} = - r^2$, and
the components of $\vec{E}$ are
\begin{equation}
E_z  \equiv  -\frac{G_{r \theta}}{r} = \frac{1}{r} \frac{\partial C_{\theta}}{\partial r}  + E^s_z \,\, , \,\, 
E_r \equiv-\frac{G_{ \theta z}}{r} = -\frac{1}{r} \frac{\partial C_{\theta}}{\partial z} ,
\label{Ez}
\end{equation}
with
\begin {equation}
C_\theta  = -rC(r,z) \, Y .
\label{defineC}
\end{equation}
The Higgs fields $\phi_i = \phi_i (r, z)$ are independent of $\theta$ and $t$.

The decomposition (\ref{effectivemag}) of $C_{\mu}$ takes the form:
\begin {equation}
C_0 = 0,\, \, \,
\vec{C}  = {\vec{C}}^D + \vec{c} \equiv 
(C^D (r,z) + c (r,z)){\hat{e}}_\theta Y, 
\label{Cnew}
\end{equation}
where
\begin {equation}
C^D (r, z) =  \frac{1}{4 \pi r} \left [ \frac{ z - R/2}{\sqrt{ r^2 +  (z - R/2)^2}} - \frac{z + R/2}{\sqrt{r^2 + (z + R/2)^2}} \right ]\,\,\frac{2 \pi}{g_m}\,\,  
\label{CsupD}
\end{equation}
is the perturbative potential of the quark sources generated by the 
Dirac string (\ref{Es}), and $c (r, z)$
is the non-perturbative potential generated by the induced
currents (\ref{cmueqofmotion}).

The color electric field (\ref{Eagain}) becomes the sum of
a Coulomb field $ {\vec{E}}_C$ 
and a non-perturbative contribution  ${\vec{E}}_{class}$:

\begin{equation}
\vec{E} = {\vec{E}}_C (\vec{x}, R) + \vec{E}_{class} (\vec{x}, R)\,\, ,
\label{Ereplaced}
\end{equation}

where
\begin{equation}
\vec{E}_C = \frac{1}{4 \pi} \left (\frac{\vec{x} -\vec{x}_1}{|\vec{x} - \vec{x}_1|^3} - \frac{\vec{x}- \vec{x}_2}{|\vec{x} - \vec{x}_2|^3} \right)\,\,\frac{2\pi}{g_m}Y,\,\, \,\,  \,\, \vec{E}_{class}  (\tilde{x}, R)= -\vec{\nabla}\times \vec{c}.
\label{Ecoulomb}
\end{equation}

At large distances ${\vec{E}}_{class} $ screens the Coulomb field  
while the Higgs fields approach their vacuum values $\phi_{i0}$, so that
the boundary conditions  are:
\begin {equation}
\vec{c}  \rightarrow - \vec{C^D},\,\,\,\,\phi_i  \rightarrow \phi_{i 0},~~~r \rightarrow \infty~~ or ~~z \rightarrow \infty.00
\label{newBC}
\end{equation}

(\ref{sumS}),  evaluated for static quarks
yields the heavy quark potential 
as the sum of a Coulomb potential $V^C (R)$ and  a
non-perturbative potential $V^{class} (R)$, where:
\begin{equation}
-S^{MAX} =T~ V^C (R),~~~~~V^C (R) = - 2 tr (\frac{2 \pi Y}{g_m})^2 (\frac{1}{4 \pi R}); ~~~
\label {VofR}  
\end{equation}
\begin {equation}
 - S^{class}(\tilde{x})    = T~V^{class}(R),~~~~~V^{class} (R) =  \int d \vec{x} ~T_{0 0} (\vec{x}, R),
\label{heavyV}
\end{equation}
and $ T_{00}(\vec{x}, R)$ is the non-perturbative contribution to the energy density: 
\begin{equation}
T_{00}(\vec{x},R) = 2 tr  [\frac{\vec{E}_{class} (\vec{x},R)^2}{2} +\frac{ g_m^2 \vec{ C} ( \vec{x})^2 [i Y, \phi_i (\vec{x})]^2}{2}] 
+2 tr[  \frac{(\vec \nabla \phi_i (\vec{x}))^2}{2}  ] + V(\phi_i) .
\label{T00R}
\end{equation}

(\ref{Sgagain}), evaluated for static quarks, becomes

\begin {equation}
\int dx [  2 tr [\frac{1}{2} G^s_{\mu \nu} G^{\mu \nu}_{class}]
 \equiv T~ W(R), 
\label{WT}
\end{equation}

where  
 
\begin {equation}
W(R)~=~\int_{-R/2}^{R/2} d z \,  2tr [- \frac{ 2 \pi}{g_m} Y \,
{\hat{e}}_z \cdot  \vec{E}_{class}(r=0, z, R)],
\label{whatisW}\end{equation}
 the work required to separate
a quark-antiquark pair a distance $R$ in the field 
$\vec {E}_{class} (\vec{x},R )$.

(\ref{Wfield}), evaluated for static
quarks, becomes the relation
\begin {eqnarray}
(S_g (\tilde{x}) - S_\phi (\tilde{x}) )
= - T~ \int d \vec{x} \frac {T_{\theta \theta} (\vec{x},R)}{r^2}  = - ~2\pi\, T \int  dz ~r \, dr\,\frac{ T_{\theta \theta} (r, z, R)}{r^2},
\label{sphisg}
\end{eqnarray}

\noindent where $\frac{T_{\theta \theta} (\vec{x},R)}{r^2}$ is the $\theta \theta$ component of the stress tensor for finite values of $R$: 
\begin{equation}
\frac{T_{\theta\theta} (\vec{x},R)}{r^2} = 2 tr  [\frac{\vec{E}_{class} (\vec{x},R)^2}{2} +\frac{ g_m^2 \vec{C}( \vec{x})^2 [i Y, \phi_i (\vec{x})]^2}{2}] 
- (2 tr [ \frac{(\vec \nabla \phi_i (\vec{x}))^2}{2}]  + V(\phi_i) )
\label{Tsum}.
\end{equation}

\noindent (\ref{Tsum}) expresses $T_{\theta \theta}$
as the difference between a
repulsive gauge contribution and
the attractive Higgs contribution 
produced by the circulating magnetic currents generated by
the Higgs condensate. 

Using (\ref{VofR}), (\ref{WT})  and (\ref{sphisg})   the 
decomposition (\ref{newSclass}) of
$S^{class}(\tilde{x})$ 
becomes a  corresponding decomposition  
of the heavy quark potential:
\begin {equation}
V^{class} (R) =   W(R) - \int  d \vec{x}\frac{ T_{\theta \theta} (\vec{x},R)}{r^2}.
\label{VWT}
\end{equation}


\subsubsection {Physical Interpretation of the Representation (\ref{VWT}) of  $V^{class} (R) $}
\label{physical}

The quantity 
$\int  dz~ r \, dr\,\frac{ T_{\theta \theta} (r, z, R)}{r^2}$
is the total torque  ${\cal{T} (R) }$ acting across any $(r,~z)$ plane bounded by the axis of the flux tube. (See Fig. \ref{fluxtube}.)
Then ${\cal{T} (R)} \Delta \theta$
is  the  work required  to remove the
field energy  in a sector of the flux tube of angular width $\Delta \theta$ between two (r, ~z) planes
 while maintaining the quark-antiquark 
separation $R$. Since the torque is independent of $\theta$,
the work required to remove all the field energy while maintaining the quark-antiquark 
separation $R$ is just $2 \pi  {\cal{T}(R) }
= \int d \vec{x} \frac{T_{\theta \theta} (\vec{x},R)}{r^2} $. 
If ${\cal{T} (R)} > 0 $ (net repulsion)  it takes work to remove the field energy.

The heavy quark potential $V^{class} (R)$ is the energy available for 
doing work when a separated quark-antiquark pair
come together.
Eq. (\ref{VWT}) expresses $V^{class} (R)$ as the difference between $W(R)$,
the work necessary to separate the pair in the fixed field
$\vec{E}_{class} (\vec{x}, R)$, and $ 2 \pi \cal{T} (R)$, the work necessary to
remove the field energy created by their separation.

\subsection{Limit $R \rightarrow \infty \,\,$ ($R~ \gg ~\frac{1}{M}$ )}
\label{Infinity}

As $R \rightarrow \infty$,
\begin{equation}
C^D (r, z) \rightarrow  -\frac{1}{g_m r}, \,\,\,\,c(r,z) \rightarrow C(r), 
\,\, \,\,\phi_i (r, z) \rightarrow \phi_i (r),
\label{limCdirac} 
\end{equation}
\begin{equation}
\vec{E}_{class} (\vec{x} , R ) \rightarrow \vec{E} (r)
= -\frac{1}{r}\frac{d(rC(r))}{dr} Y {\hat e}_z ,
\label{Enew}
\end{equation}
\begin {eqnarray}
\label{WR}
W(R)& \rightarrow& 2 tr \left[ -\frac{2 \pi}{g_m} Y \hat{e}_z \cdot \vec{E} (r=0) \right ]  ~R,\\  \nonumber
T_{0 0} (\vec{x},R) &\rightarrow& T_{0 0} (r),  \\  \nonumber
T_{\theta \theta} (\vec{x},R)& \rightarrow &T_{\theta \theta} (r),  \label{VclassR} \\ 
V^{class}(R)  =\int d \vec{x} T_{00} (\vec{x}, R) & \rightarrow &  \int_0^\infty 2 \pi r dr T_{0 0} (r) ~ R ~=~\sigma R, 
\\
2 \pi {\cal{T}}(R)~=~\int  d \vec{x}~ \frac{ T_{\theta \theta} (\vec{x}, R)}{r^2} & \rightarrow &   \int_0^\infty 2 \pi r dr \frac{ T_{\theta \theta} (r)}{r^2} ~R~=~2 \pi \tau R; 
\label{tauR}
\end{eqnarray}

\noindent where $\tau$ is the torque per unit length (\ref{tau}), and

\begin{eqnarray}
\label{Tftprime}
T_{0 0} (r)& =&
2 tr \left [ \frac{1}{2} \vec{E}^2 (r) + g^2_m(C(r)-\frac{1}{g_m r})^2 [iY, \phi_i]^2 \right]
+2 tr\left [\frac{1}{2} (\frac{d \phi_i (r)}{dr})^2 \right ] + V(\phi_i) , \nonumber \\ 
\frac{T_{\theta\theta} (r) }{r^2}&= &
2 tr \left [ \frac{1}{2} \vec{E}^2 (r) + g^2_m(C(r)-\frac{1} {g_m r})^2[iY, \phi_i]^2 \right]
-2 tr\left [\frac{1}{2} (\frac{d \phi_i (r)}{dr})^2 \right ] - V(\phi_i) . \nonumber  \\
\end{eqnarray}

Taking the large $R$ limit of   (\ref{VWT}),  using (\ref{WR}), (\ref{VclassR}) and (\ref{tauR}) yields  Eq (\ref{sigmaWprime}),  as stated in the Introduction.  Eq (\ref{sigmaWprime}) links the string tension $\sigma$ to the field $\vec{E} (r=0)$ on the axis of an infinite flux tube via the parameter $\tau$, and has the physical interpretation discussed in the Introduction and in the previous section.

 Using the fact that $\sigma$ is the long distance force on a quark,

\begin {equation}
\sigma = 2 tr \left [ -\frac{2 \pi}{g_m} Y \vec{E}_{class} (r=0, z= \pm \frac{R}{2}; R) \right ] \cdot \hat{e}_z,~~~ R \gg \frac{1}{M},
\label{second sigma}
\end{equation}

\noindent we can write (\ref{sigmaWprime}) in an alternate form: 

 \begin{equation}
 2 tr \left [ -\frac{2 \pi}{g_m} Y \vec{E}_{class} (r=0, z= \pm \frac{R}{2}; R) \right ] = 2 tr [-\frac{2 \pi}{g_m} Y \vec{E}(r=0) ] - 2 \pi \tau \hat{e}_z,~~~ R \gg \frac{1}{M}.
 \label{chromofield}
 \end{equation}

\noindent  The field  on the axis of an infinite flux tube
 $\vec{E}( r=0)$ is  equal to the field of a quark and antiquark,
  $\vec{E}_{class} (r=0, z; R)$, evaluated in the central region $|z| \ll \frac{R}{2}$, far from the positions of the quarks. Consequently, (\ref{chromofield}) has the equivalent form:
  
  \begin {equation}
  2 \pi \tau \vec{e}_z =    2 tr \left [ -\frac{2 \pi}{g_m} Y \vec{E}_{class} (r=0, z= \pm \frac{R}{2}; R) \right ]  -  2 tr \left [ -\frac{2 \pi}{g_m} Y \vec{E}_{class} (r=0, z; R) \right ],~~~~|z| \ll \frac{R}{2}, ~~~R \gg \frac{1}{M}.
  \label{diffEclass}
  \end{equation}
  The torque per unit length $\tau$ thus determines the difference between the value of the field $\vec{E}_{class}$  at the positions of the quarks and its value midway between them.
\noindent ((\ref{diffEclass}) is an equivalent characterization of the parameter  $\tau$.)

\section{A New Constraint on the QCD Flux Tube}
\label{Speculation}
We now  assume that the value $\tau=0$  characterizes the QCD flux tube and examine the consequences of this constraint. 

If $\tau=0$  the string tension  equals the color charge $\frac{2 \pi}{g_m} Y$ of the quark multiplied by the  field $\vec{E} (r=0)$  on the axis of an infinite $Z_N$ flux tube  (Eq(\ref{newsigma}));  i. e.,  the force on a quark in the field of the 'string' connecting the pair.

Further,  (\ref{diffEclass}) becomes the equality


\begin {equation}
\vec{E}_{class} (r=0, z= \pm \frac{R}{2}; R) = \vec{E}_{class} (r=0, z; R),~~~ |z| \ll \frac{R}{2}, ~~~~~ R \gg \frac{1}{M},
\label{new field}
\end{equation}
\noindent so that the field  at the positions  of the quarks equals the field  in the middle of the flux tube.

A non vanishing value of $\tau$ necessitates a variation of
$\vec{E}_{class} (r=0, z; R)$ along the line connecting the pair.
The condition $\tau = 0$  allows this field 
to remain constant for all $z$ including points close to the positions of the quarks. (Expressed in this way one might speculate that the  condition $\tau=0$ imposed on the effective field theory reflects a flicker of  the short distance asymptotic freedom of the fundamental theory visible in the effective field theory. )

\subsection {The Action of the Effective String Theory}
\label{action}

 Poincare invariance implies that  the action of the effective string theory obtained from effective field theory $ S_{eff} (\tilde{x}) = S^{class} (\tilde{x})|_{\frac{1}{M} = 0}~=~S_{NG} (\tilde{x})$ for any value of $\tau $ (Eq (\ref{newaction})). We will now show that under the condition $\tau=0$, $ S^{class} (\tilde{x})|_{\frac{1}{M} = 0}$ has a representation in terms of  the chromodynamic fields of magnetic $SU(N)$ gauge theory.  This will  give a field theory interpretation of effective string theory. 

For long straight strings  (\ref{sphisg}) and (\ref{tauR}) show that the term  linear in $R$ in $S_g (\tilde{x}) - S_\phi (\tilde{x})$ is proportional to $\tau$. Hence for curved strings, by Poincare symmetry the term having the Nambu-Goto form in  $S_g(\tilde{x}) - S_\phi(\tilde{x})$  is also proportional to $\tau$ \cite{sergei}.  Thus  if $\tau=0$, 
 $S_g(\tilde{x}) - S_\phi(\tilde{x})$ does not contain a term proportional to the Nambu-Goto action, and 
 can be neglected on the right hand side of Eq (\ref{newSclass}) for $S^{class} (\tilde{x})$;   its contribution to $S_{eff} (\tilde{x})$ generates terms in the ground state heavy quark potential that fall off faster than $\frac{1}{R^5}$ \cite{aharony1, aharony2}. Then (\ref{newSclass}) takes the form
             
\begin{eqnarray}
\label{newSclassprime}
 S^{class}(\tilde{x}) |_{\frac{1}{M} =0}~&=&~-
 \int dx 2 tr  [\frac{1}{2} G_{\mu \nu}^s G_{class}^{\mu \nu}]\\ \nonumber
& =&
\frac{1}{4} \int d \tau \int d \sigma \sqrt{-g}  \epsilon_{\mu \nu \lambda \alpha}\, 2 tr \left( \frac{2 \pi}{g_m} Y  G^{\mu \nu}_{class} (x, \, \tilde{x}) |_{x^\mu
= \tilde{x}^\mu (\sigma, \tau) }  \right )\,\, t^{\lambda \alpha}( \sigma,  \tau),
\end{eqnarray}
an integral of the field 
tensor $G_{class}^{\mu \nu} (x, \tilde{x})$ evaluated 
on the vortex sheet $x^\mu = \tilde{x}^\mu (\sigma, \tau)$. 
Eq (\ref{newSclassprime}) gives
the Nambu-Goto action a representation solely in terms of
the chromodynamic fields 
of the 4-dimensional effective field theory.

Writing (\ref{newSclassprime}) in a parameterization where 
$ \frac{\partial \tilde{x}^0}{\partial \sigma} |_{\tau =0} $, using (\ref {Wtau}) and (\ref{F})  
gives 

\begin{equation}
S_{NG} (\tilde{x}) 
 = - \int  d \tau \int d \sigma \frac {  \partial \vec {\tilde{x}}}{\partial \sigma} \cdot  \vec {F}_{class}  (\sigma, \tau, \tilde{x}) ( \frac{\partial \tilde{x}^0}{\partial \tau}),
\label{Wtauprime}
\end{equation}
where $\vec{F}_{class} (\sigma, \tau, \tilde{x})$ is the chromodynamic force (\ref{F}) acting along the string. (\ref{Wtauprime}) is the representation of the Nambu-Goto action in terms of  fields and is the generalization of the relation (\ref{newsigma}) to curved vortex sheets.

\subsection{The Relation Between Fields and Surfaces}
\label{fields} 

For a curved vortex sheet
Lorentz invariance and reparameterization invariance imply that 
 $\epsilon_{\mu \nu \lambda \alpha} G^{\lambda \alpha}_{class} (x, \tilde{x})|_ {x^\mu = \tilde{x}^\mu (\sigma, \tau)} $ 
must be proportional to  the
tensor (\ref{tab})  describing the 
orientation of the world sheet $\tilde{x}^\mu (\sigma, \tau)$:

\begin{equation}
2tr\left (\frac{2 \pi}{g_m} Y\frac{1}{2} \epsilon_{\mu \nu \lambda \alpha}  G^{\lambda \alpha}_{class} (x, \tilde{x})|_ { x^\mu =\tilde{x}^\mu (\sigma, \tau)} \right) = \,\sigma\, t_{\mu \nu} (\sigma, \tau). 
\label{curved}
\end{equation}
Consistency of (\ref{curved}) evaluated
for a long straight vortex with  (\ref{newsigma}) fixes  the string tension
$\sigma$ as the coefficient of $t_{\mu \nu}$.
Taking into account non-leading terms in $\frac{1}{M}$  would introduce 
higher dimensional tensors and new parameters on the right
hand side of (\ref{curved}). \footnotemark

 \footnotetext [3]{Relations between fields and surfaces,
postulated  on the basis of symmetry, with account taken of non-leading 
terms and limited to the positions of the quarks, have been used
to calculate  heavy quark potentials \cite{Noraetal}.}

Therefore, to leading order in $\frac{1}{M}$ 
the values of the  chromodynamic fields  $G^{\mu \nu}_{class}$ on the vortex sheet
are determined in terms of the string tension $\sigma$ and the geometry of the vortex sheet.
Using (\ref{curved}) in  (\ref {newSclassprime}) and the normalization $t^{\mu \nu} t_{\mu \nu} = -2$ 
of the surface tensor 
yields the Nambu-Goto action directly.

Expressing (\ref{curved}) in terms of the color electric and magnetic components (\ref{EBclass}) of $G^{\mu \nu}_{class}$ gives the  values of the
fields $\vec{E}_{class}$ and $\vec{B}_{class}$ on the
vortex sheet:

\begin {equation}
2 tr [-\frac{2 \pi}{g_m} Y E^k_{class}]|_{x^\mu = \tilde{x}^\mu} = \sigma t^{0k} (\sigma, \tau),~~~2 tr[ \frac{2 \pi}{g_m} Y B^k_{class}]|_{x^\mu = \tilde{x}^\mu} =\frac{ \sigma}{2} \epsilon_{klm}  t^{lm}  (\sigma, \tau).
\label{sigmafluxtube}
\end{equation}

\noindent We choose $\tau = t,~\sigma = z$, and a parameterization 
$\tilde{x}^\mu (z, t)$   of the vortex sheet in terms of the two transverse fluctuations
$\vec{x}^i_\perp (z, t),~~i = 1,2$:
 
 \begin {equation}
  \tilde{x}^\mu (z, t) = x^\mu (t,~z,~ \vec{x}^1_\perp (z, t),\,\vec{x}^2_\perp (z, t)).
  \label{xmuperp}
  \end{equation}
  
The color fields  evaluated on the vortex sheet are corresponding functions of $z$ and $t$:

  \begin{equation}
  \vec{E}^i_{class} (x, \tilde{x})|_{x^\mu \equiv \tilde{x}^\mu (z, t)} ~ \equiv ~ \vec{E}^i_{class} (z, t),~~  \vec{B}^i_{class} (x, \tilde{x})|_{x^\mu = \tilde{x}^\mu (z, t)} ~ \equiv ~ \vec{B}^i_{class} (z, t),
  \label{EBcorresponding}
  \end{equation}

  Eq.
(\ref{sigmafluxtube}), with use of (\ref{tab}) and (\ref{xmuperp})   determines the fields
$\vec{E}_{class} (z, t) $ and $\vec{B}_{class} (z, t )$  in terms 
of the  transverse fluctuations  $\vec{x}^i_\perp (z, t)$ for $-R/2~\le~ z~ \le~ R/2$:

  \begin {eqnarray}
    \label{newEandBclass}
  2 tr [ -\frac{2 \pi Y}{g_m} \vec{E}^i_{class} (z,t) ] &=& 
  \frac{\sigma}{\sqrt {-g}} 
  \frac{\partial \vec{x}^i_\perp}{\partial z},\\  \nonumber
    2 tr [  \frac{2 \pi Y}{g_m} \vec{B}^i_{class} (z,t) ] &=& \frac{\sigma}{\sqrt {-g}}( \hat{e}_z \times \frac{\partial \vec{x}_\perp}{\partial t} )^i,\\ \nonumber
      2 tr [ -\frac{2 \pi Y}{g_m} \vec{B}^z_{class} (z,t) ] &=& \frac{\sigma}{2 \sqrt {-g}} (\frac{\partial \vec{x}_\perp}{\partial t} \times\frac{\partial \vec{x}_\perp}{\partial z} ) \cdot\hat{e}_z,\\ \nonumber
  2 tr [ -\frac{2 \pi Y}{g_m} \vec{E}^z_{class} (z,t) ] &=& 
  \frac{\sigma}{\sqrt {-g}}.\\  \nonumber 
      \end{eqnarray}

The Wilson loop $W(\Gamma)$ written in the
parameterization (\ref{xmuperp}) is

\begin {equation}
W(\Gamma)=  \int {\cal{D}} \vec{x}^1_\perp~{\cal{D}} \vec{x}^2_\perp~ exp [-i \sigma  \int dt \int_{-R/2}^{R/2} dz \sqrt{-g(\tilde{x}^\mu (z,t))}~]. 
\label {allnew}
\end{equation}


\noindent If $\tau=0$ the Nambu-Goto action has the representation (\ref{Wtauprime}), so that
the Wilson loop (\ref{allnew}) can be used 
along with the relations (\ref{newEandBclass}) to calculate 
correlation functions of the fields $\vec{E}_{class} (z, t)$
and $\vec{B}_{class} (z, t)$ and physical quantities dependent 
on them.

 We now describe the picture that results  from the condition $\tau=0$ in a particular model.

\section{SU(N) Vortices in a Particular Model }
\label{specific}

The effective Lagrangian in the model \cite{BBZ:1990}  has the form (\ref{Leff})
with 3 scalar Higgs fields and a Higgs potential $V(\phi_ i)$
generated from one loop contributions to the
scalar 2 point and 4 point functions in effective $SU(N)$ magnetic gauge theory: 

\begin{equation}
V(\phi_i) = \mu^2 N  \sum_i  2 tr [\phi_i^2]  + \frac {4 N \lambda}{3} \left ( tr  (\sum_{i\,j}  \phi_i^2 \phi_j^2) + \frac{1}{N}(tr ( \sum_i  \phi_i^2) )^2 + \frac{2}{N} \sum_{i\,j}( tr \phi_i \phi_j )^2 \right),
\label{Vphii}
\end{equation}
where the parameter $\mu^2$ has dimensions of mass squared and  $\lambda$ is dimensionless.

In the confining vacuum the Higgs condensate $\phi_{i \, 0}$ has the color
structure:
\begin {equation} 
\phi_{10} = \phi_0 J_x,\,\,\,\,\, \phi_{20} = \phi_0 J_y,\,\,\,\,\, \phi_{30} = \phi_0 J_z,
\label{phii0}
\end{equation}
where $J_x,\,\,J_y,$ and $J_z$ are the three generators of the N-dimensional irreducible representation of the 3 dimensional rotation group corresponding to angular momentum $J = \frac{N-1}{2}$. 
Since any  matrix which commutes with all three generators $J_i$ must be a multiple of the unit matrix, there is no SU(N) transformation which leaves all three $\phi_i$ invariant and the dual $\frac{SU(N)}{Z_N}$
gauge symmetry is completely broken.

 The  Higgs potential has  an absolute minimum at $\phi_i = \phi_{i0}$ with $\phi_0^2 = - \frac{9 \mu^2}{8 (N^2 -1 ) \lambda}$.
The difference $\epsilon_V$ between the energy density of the symmetry breaking vacuum $\phi_i = \phi_{i0}$ and the perturbative vacuum $\phi_i =0$ is the minimum value  $V(\phi_{i 0})$ of the Higgs potential:
\begin{equation}
\epsilon_V = V(\phi_{i0}) = \frac{-\lambda}{9} \left ((N(N^2 -1) \phi_0^2 \right)^2.
\label{epsV}
\end{equation}

\subsection {The Classical Action for SU(3) Vortices}
\label{SU(3)}

For SU(3) 
\begin {equation}
J_x = \lambda_7,\,\,\,J_y = -\lambda_5,\,\,\,J_z =  \lambda_2,\,\,\,Y = \frac{\lambda_8}{\sqrt{3}} \,\, ,  
\label{Jx}
\end{equation}
and the vector mass  (\ref{explicitM2}) has the value
\begin {equation}
M = \sqrt 6 g_m \phi_0.
\label{M}
\end{equation}

We make the following  ansatz for the classical solution:  
\begin{eqnarray}
\label{newphi}
\phi_1 & = & \phi_1(x, \tilde{x}) \frac{(\lambda_7 - i \lambda_6)}{2} + \phi_1^* (x, \tilde{x}) \frac{(\lambda_7 + i \lambda_6)}{2},\\  \nonumber
\phi_2 & = &\phi_2(x, \tilde{x}) \frac{(-\lambda_5 - i \lambda_4)}{2} + \phi_2^* (x, \tilde{x}) \frac{(-\lambda_5 + i \lambda_4)}{2},\\  \nonumber
\phi_3 & = &\phi_3 (x, \tilde{x}) \lambda_2,\\  
\label{Cmu2}
C_\mu  & = &C_\mu (x) Y  = (C^D_\mu (x, \tilde{x}) + c_\mu (x, \tilde{x})) Y, \\  \nonumber
G_{\mu \nu \,\, class} &= &(\partial_\mu c_\nu (x) - \partial_\nu c_\mu (x))Y \equiv  G_{\mu \nu  \,\, class} (x, \tilde{x})Y.\\   \nonumber
\end{eqnarray}
There are two other solutions, physically equivalent to  (\ref{newphi}),
related by gauge transformations 
taking $Y \rightarrow -\frac{Y+\lambda_3}{2}$ or 
$Y \rightarrow \frac{\lambda_3-Y}{2}$, 
corresponding to the other two quark colors  \cite{konishi}.

The commutation relations 
\begin {equation}
[Y, \lambda_7 - i \lambda_6] = \lambda_7 - i \lambda_6, \,\,\, [ Y, -\lambda_5 - i \lambda_4] = - (-\lambda_5 - i \lambda_4),\,\, [Y, \lambda_2] =0
\label{commutation}
\end{equation}
yield
\begin {eqnarray}
\label{calDphi1}
{\cal{D}}_\mu \phi_1 =
(\partial_\mu - i g_m C_\mu (x)) \phi_1(x)
\frac{(\lambda_7 - i \lambda_6)}{2} + 
(\partial_\mu + i g_m C_\mu (x)) \phi^*_1(x)
\frac{(\lambda_7 + i \lambda_6)}{2},\\ \nonumber
{\cal{D}}_\mu \phi_2 =
(\partial_\mu + i g_m C_\mu (x)) \phi_2(x)
\frac{(-\lambda_5 - i \lambda_4)}{2} +
(\partial_\mu - i g_m C_\mu (x)) \phi^*_2(x)
\frac{(-\lambda_5 + i \lambda_4)}{2},\\ \nonumber
\end{eqnarray}

\noindent so that the Higgs fields $\phi_1$ and $ \phi_2$ carry $Y$ charge $\pm 1$ 
and that $\phi_3$ carries  $Y$ charge $0$.

The consistency requirement that  the magnetic 
current density (\ref{cmueqofmotion})
be proportional to $Y$ forces
\begin {equation}
\phi_1 (x,~\tilde{x})\,=\,\phi_2^* (x,~\tilde{x}) \,\equiv\, \phi (x,~\tilde{x}),
\label{forcedphi}
\end{equation}
and yields
\begin {equation}
j^\mu  
= 6g_m (\frac{\phi^* (x)D^\mu \phi (x) - \phi (x)(D^\mu \phi (x))^*}{2 i} )Y \,\,,
\label{6Y}
\end{equation}
where
\begin {equation}
D_\mu \phi (x) \equiv (\partial_\mu - i g_m C_\mu (x)) \phi(x).
\label{Dmuphi}
\end{equation}

Using the color ansatz  (\ref{Cmu}) and (\ref{newphi}) in (\ref{delta})  and  
(\ref{Vphii}), making use of (\ref{forcedphi}) and subtracting off the 
vacuum energy density $\epsilon_V$ gives $S^{class} (\tilde{x})$ the form:
\begin {equation}
S^{class}(\tilde{x})  =  \int dx \left[\frac{4}{3} (-\frac{1}{4} G_{\mu \nu \,\, class} (x) G^{\mu \nu}_{class} (x))+ 4 (D_\mu \phi(x)) (D^\mu \phi (x))^* + 2 \partial_\mu \phi_3 (x)~ \partial^\mu \phi_3 (x) - V(\phi, \phi_3) \right] \,\, ,
\label{newLeff}
\end{equation}
where
\begin {equation}
V(\phi,\,\,\phi_3) = \frac {22 \lambda}{3} (2 (|\phi|^2  - \phi_0^2)^2 + (\phi_3^2 - \phi_0^2)^2 ) + \frac {14 \lambda}{3} (2 |\phi|^2 + \phi_3^2 - 3 \phi_0^2)^2.
\label{newV}
\end{equation}

The corresponding field equations are
\begin{equation}
\partial^\mu G_{\mu \nu}^{class} (x,~\tilde{x}) =  \partial^\mu \partial_\mu c_\nu - \partial_\nu \partial^\mu c_\mu  =  6 g_m (\frac {\phi^* \partial_\nu \phi - \phi \partial_\nu \phi^*}{2 i} - g_m C_\nu \phi^* \phi),
\label{dg}
\end{equation}
and
\begin {equation}
-D_\mu D^{\mu*} \phi(x) = \frac{1}{4} \frac{ \delta V } {\delta \phi^*(x)}, ~ -\partial_\mu \partial^\mu \phi_3(x) = \frac{1}{2} \frac{\delta V}{\delta \phi_3(x)}.
\label{newclassicaleqs}
\end{equation}


At large distances the Higgs fields are a gauge transformation of the vacuum solution (\ref{phii0}). 
With an appropriate gauge transformation $(\ref{abgt})$ the field $\phi (x, \tilde{x})$
can be made real. The boundary conditions at large distances are
then

\begin {equation}
\phi (x, \tilde{x}) \rightarrow \phi_0,\,\,\,\,\,\phi_3 (x, \tilde{x}) \rightarrow \phi_0,\,\,\,\,\, c_\mu(x, \tilde{x} )\rightarrow - C_\mu^D (x, \tilde{x}) .
\label{gaugetrans}
\end{equation}

On the vortex sheet $\tilde{x}^\mu (\sigma, \tau)$ where $C^D_\mu (x, \tilde {x})$ is singular  the boundary conditions are:

\begin{equation}
\phi(x, \tilde{x})|_{x^\mu = \tilde{x}^\mu (\sigma, \tau)} = 0,\,\,\,\,\,\,\phi_3 (x, \tilde{x})|_{x^\mu = \tilde{x}^\mu (\sigma, \tau)} = finite.
\label{bc0}
\end{equation}
Eqs. (\ref{dg}) and (\ref{newclassicaleqs}) were solved
for the flat vortex sheet (\ref{Es}), and the resulting static 
heavy quark potential  $V^{class} (R)$  determined in \cite{BBZ:1991}.

\subsection{Static Flux Tube Solutions }
\label{Static}

For the infinite $Z_3$ flux tube the vector potential $\vec C$ 
 has the form 
(\ref{Cft})
with $Y = \frac{\lambda_8}{\sqrt{3}}$ , and  
the Higgs fields (\ref{phiofomega})
 are obtained by making the gauge transformation $\Omega (\theta) = exp[i \theta Y]$ to the color ansatz
(\ref{newphi}) with $\phi_1 (x, \tilde{x}) = \phi^*_2 (x, \tilde{x}) = \phi (r), ~~  \phi_3 (x, \tilde{x}) = \phi_3 (r); $

\begin{eqnarray}
\label{phistransformed}
& exp[-i \theta Y] \phi_1 \, exp [i \theta Y]& = \phi (r)exp[-i  \theta] \frac{\lambda_7- i \lambda_6 }{2}+ \phi (r) exp[i \theta]  \frac{\lambda_7+ i \lambda_6 }{2}, \\  \nonumber
& exp[-i \theta Y] \phi_2 \, exp [i \theta Y]& = \phi (r)exp[i  \theta] \frac{ -\lambda_5 -i \lambda_4 }{2}+ \phi (r) exp[-i \theta]  \frac{-\lambda_5+i \lambda_4 }{2}, \\  \nonumber
& exp[-i \theta Y]\,\phi_3 \, exp [i \theta Y]& = \phi_3 (r) \lambda_2.\  \nonumber
\end{eqnarray}
Then the Higgs fields in the infinite flux tube  (\ref{phistransformed}) have the color structure (\ref{newphi}) with
 \begin {equation}
\phi_1(x, \tilde{x}) = \phi^*_2 (x, \tilde{x}) = \phi (x, \tilde{x}) = \phi(r) exp (-i \theta),\,\,\,\phi_3 (x, \tilde{x}) = \phi_3 (r).
\label{classB}
\end{equation}

 (\ref{phistransformed}) gives the specific form of (\ref{phiofomega}) for the $SU(3)$ flux tube in the gauge where $C_\mu = C_\mu (x) Y$, with $Y = \frac{\lambda_8}{\sqrt{3}}$. 
Replacing $Y$ by $-\frac{Y+\lambda_3}{2}$ or by $\frac{\lambda_3-Y}{2}$ 
on the right hand side of (\ref{Cmu}) 
(corresponding to the other two quark colors)
yields three physically equivalent vortices, each carrying one unit of 
$Z_3$ flux, related by $SU(3)$ gauge transformations.

We rescale the flux tube fields, choosing the flux tube 
radius $\frac{1}{M}$ as the scale of length, making the replacement

$$ r \rightarrow r/M,\,\, C(r) \rightarrow \frac{M C(r)}{g_m}, \,\,
\phi (r) \rightarrow \phi_0 \phi, \,\, \phi_3 (r) \rightarrow \phi_0 \phi_3, \,\,\, \phi_0^2 = \frac {M^2}{6 g_m^2} ,$$ and 
define a rescaled Higgs potential $W( \phi, \phi_3)$: 
\begin {eqnarray}
\label{Wnew}
W( \phi,  \phi_3) &\equiv &   \frac{1}{96} \frac{V(\phi_0 \phi,  \phi_0 \phi_3)}{M^4 g_m^2 } \nonumber \\
&=& \frac{\kappa^2}{200} \left ( 11[ 2 (\phi^2 -1)^2 + (\phi_3^2 -1 )^2 ] + 7[2 (\phi^2 -1) + (\phi_3^2 -1 ) ]^2 \right ) \nonumber \\
&=&  \kappa^2 \left (   \frac{ (\phi^2 -1)^2}{4}  +9 \frac{ (\phi_3^2 -1 )^2}{100} - 7  \frac { (\phi_3^2 -1) (1 - \phi^2)}{50} \right ), \nonumber \\
\end{eqnarray}
with
\begin {equation}
\kappa^2 \equiv  \frac{25}{9} \frac {\lambda}{g_m^2}.
\label{kappa2}
\end{equation}

\noindent Note that with $\phi_3 (r)$ replaced by $1$ in (\ref{Wnew}),
$W (\phi, \phi_3)$ becomes $\frac{\kappa^2}{4} (\phi^2 -1 )^2$, the Higgs potential of the Abelian Higgs model with
Landau-Ginzburg parameter $\kappa$.

The rescaled expressions  for $T_{0 0}(r)$  
and $\frac{T_{\theta \theta}(r)}{r^2}$ (\ref{Tftprime})
are:

\begin {eqnarray}
\label {Tnew}
T_{0 0} (r)& =& \frac{4}{3} \frac {M^4}{g_m^2} (\frac{1}{2} (\frac{1}{r} (\frac{d (rC)}{dr})^2 + \frac{1}{2} (C - \frac{1}{r})^2 \phi^2+ \frac{1}{2} (\frac{d \phi}{dr})^2 +\frac{1}{4} (\frac {d \phi_3}{d r} )^2 +
W (\phi, \phi_3 ), \nonumber \\
\frac{T_{\theta \theta} (r)}{r^2}& = &\frac{4}{3} \frac {M^4}{g_m^2}  (\frac{1}{2} (\frac{1}{r} (\frac{d (rC)}{dr})^2 + \frac{1}{2} (C - \frac{1}{r})^2 \phi^2- \frac{1}{2} (\frac{d \phi}{dr})^2 -\frac{1}{4} (\frac {d \phi_3}{d r} )^2 -
W (\phi, \phi_3 )) .\nonumber \\
\end{eqnarray}

The rescaled static field equations obtained from 
$T_{0 0}(r)$  are: 
\begin {equation}
\frac{d}{dr} (\frac{1}{r} \frac{d (r C)}{dr} ) = (C - \frac{1}{r}) \phi^2,
\label{neweqofmot1}
\end{equation}

\begin{equation}
- \frac{1}{r} \frac{d}{dr} (r \frac{d \phi}{dr}) + \phi (C - \frac{1}{r})^2 + \kappa^2 \phi [(\phi^2 -1 ) + \frac{7}{25} (\phi_3^2 -1) ] =0,
\label{neweqofmot2}
\end{equation}

and

\begin {equation}
-\frac{1}{r}\frac{d}{dr} (r \frac{d \phi_3}{dr}) + \frac{2\kappa^2}{25} \phi_3 [ 7(\phi^2 - 1) + 9( \phi_3^2 - 1)] =0.
\label{neweqsofmot}
\end{equation}

with boundary conditions

\begin {eqnarray}
C(r) & \rightarrow & \frac {1}{r},\,\,\,\phi(r) \rightarrow 1,\,\,\, \phi_3 (r) \rightarrow 1\,\, as\,\, r \rightarrow \infty,\\ \nonumber
C& \rightarrow & 0, \,\,\, \phi(r) \rightarrow 0,\,\,\, \phi_3 (r) \rightarrow finite\,\,\, as\,\, r\,\, \rightarrow 0.
\label{bc}
\end{eqnarray}

The numerical solution of
(\ref{neweqofmot1}), (\ref{neweqofmot2}) and (\ref{neweqsofmot}) shows 
that  $\phi (r) < 1$ and $\phi_3 (r) >1 $ everywhere; 
hence the term coupling  $\phi$ and $\phi_3$  in $W (\phi,\, \phi_3 )$ 
is attractive.  This  additional attraction  
reduces the energy of the $Z_3$ vortex  below that of the Abelian
configuration with $\phi_3 (r) = 1$, 
viewed as an unstable configuration of the non-Abelian vortex. 

Evaluation of  (\ref{Tnew}) at the classical solution yields expressions for 
the string tension $\sigma$ and  the 
torque per unit length  $\tau$ as the sum and difference, respectively, of a gauge contribution $\sigma_g (\kappa)$ and a Higgs contribution $\sigma_h (\kappa)$:


\begin {eqnarray}
\label{sigma}
\sigma& = & \int_0^\infty 2 \pi r T_{0 0} (r) dr  =  \frac{4}{3} \frac {M^2}{g_m^2}( \sigma_g (\kappa) + \sigma_h (\kappa)) \equiv \frac{4}{3} \frac {M^2}{g_m^2} \sigma(\kappa),\\
\label{balance2}
2 \pi \tau & =& \int_0^\infty 2 \pi r dr \frac{ T_{\theta \theta} (r)}{r^2} = \frac{4}{3} \frac{M^2}{g_m^2} (\sigma_g (\kappa) - \sigma_h (\kappa)), 
\end{eqnarray} 

where

\begin{equation}
\sigma_g (\kappa)   =  \int_0^\infty 2 \pi r dr (\frac{1}{2} (\frac{1}{r} (\frac{d (rC)}{dr})^2 + \frac{1}{2} (C - \frac{1}{r})^2 \phi^2), 
\label{sigmag}
\end{equation}

and

\begin{equation}
\sigma_h (\kappa) =   \int_0^\infty 2 \pi r dr ( \frac{1}{2} (\frac{d \phi}{dr})^2 +\frac{1}{4} (\frac {d \phi_3}{d r} )^2 +
W (\phi, \phi_3 )).
\label{sigmah}
\end{equation}

The condition $\tau = 0$
becomes
 $\sigma_g (\kappa)~=~\sigma_h(\kappa)$.


\subsection {Results for $T_{\theta \theta}$ in $Z_3$ Flux Tubes}
\label{Results}

\begin{figure}[htbp]
\begin{center}
\includegraphics[width=4in]{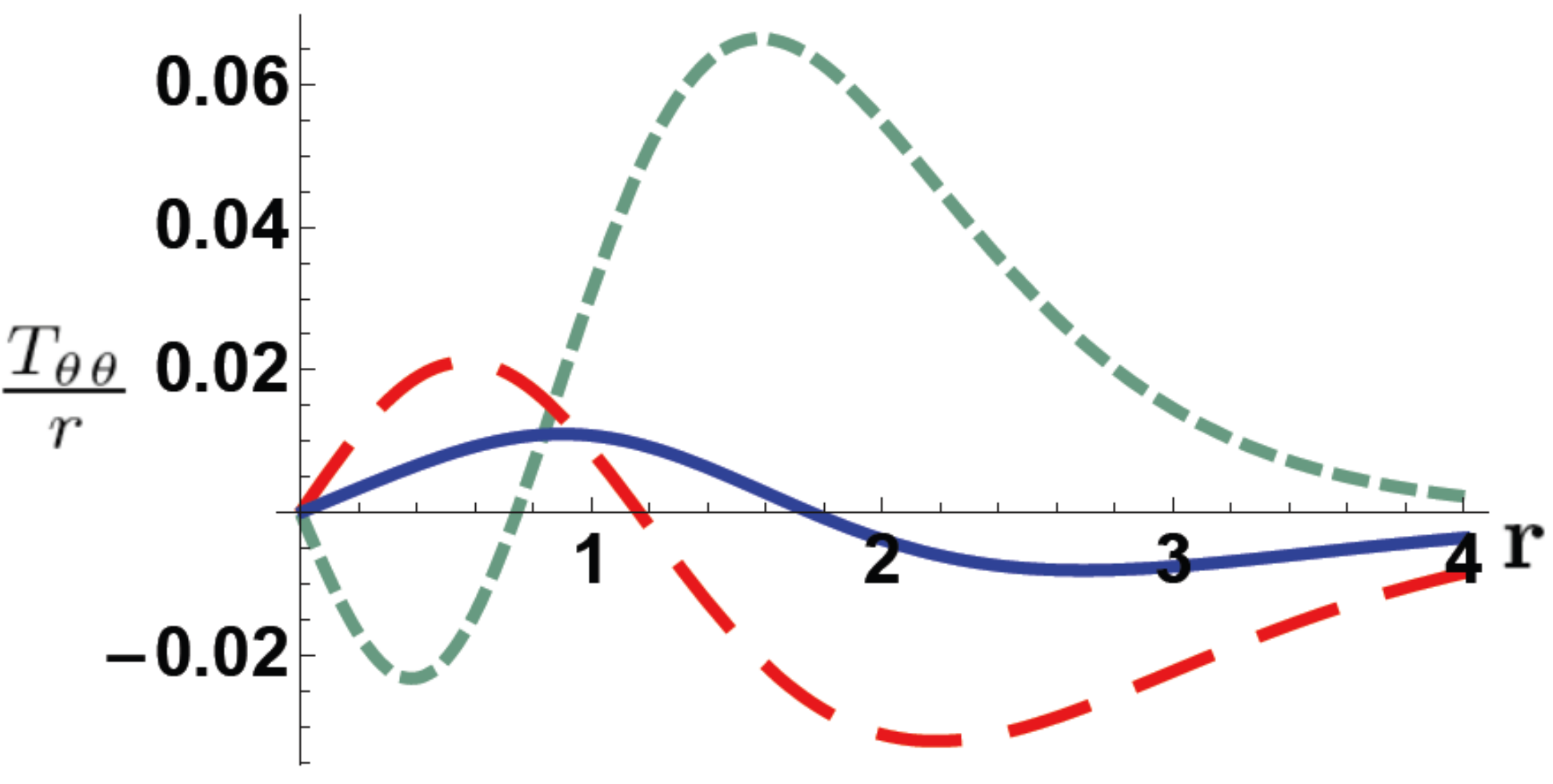}
\caption{The torque per unit area,  $T_{\theta\theta}(r)/r $.  Red, long dashed, $\kappa^2 = 0.5$;  ~~ blue, thick, $\kappa^2= 0.59$; ~~  green, short dashed, $\kappa^2 = 0.8$. }
\label{3Ttth}
\end{center}
\end{figure}

Fig. \ref{3Ttth} shows  $\frac{T_{\theta \theta}(r)}{r} \equiv r p(r)$ evaluated at
the classical solution for three values of
$\kappa^2$. The condition  (\ref{balance}), $\sigma_g (\kappa) = \sigma_h (\kappa)$, yields 
$\kappa^2 \approx 0.6$,  and $\sigma  (\kappa^2 = 0.6) \approx 3.1$. (The value of $\kappa^2 = 0.6$ lies close to the value $\kappa^2 = \frac{5}{9}$ used in \cite{BBZ:1997} in comparing calculations of heavy quark potentials in this  model with lattice simulations \cite{bali}.)  For $\kappa^2 \approx 0.6$,  the stress tensor component
$T_{\theta \theta }(r) =0 $ at $r  \equiv r^* = \frac{1.7}{M} $.   
There is repulsion  for  $r < r^*$, where $T_{\theta \theta } (r) > 0$,
and  attraction for $r > r^*$, where 
$T_{\theta \theta }(r) < 0$. It is then natural to identify 
$r^*$ as a boundary separating the repulsive interior 
of the flux tube from its attractive exterior. 


(\ref{balance}) is  also satisfied by the flux tubes of the 
Abelian-Higgs model with  $\kappa^2 = \frac{1}{2}$.
These are BPS states \cite{bogomolny}
describing an Abelian magnetic superconductor on the border between
type I and type II.  In this situation  $T_{\theta\theta} (r) $ ~ =~0
for all $r$ \cite{deVega}, so that  the profile of
$T_{\theta \theta} (r)$ does not reveal a boundary.
The difference between the non-Abelian and Abelian vortices is caused by 
the additional attractive interaction among the octet of scalar
particles which breaks the supersymmetry \cite{fayet} giving 
rise to the BPS vortex of the Abelian Higgs model. For $\kappa^2\approx 0.6$, where $\tau=0$, this additional interaction is approximately compensated for by the additional gauge repulsion associated with the fact that $\kappa^2 > \frac{1}{2}$.

\label{Deconfined}

\section {Summary}
\label{Summary}

\subsection {Relation Between Effective Field Theory and Effective String Theory}
\label{Relation}

We have started with magnetic $SU(N)$ gauge theory as an effective
field theory of the long distance heavy quark interaction
in Yang-Mills theory. At  interquark distances $R \sim \frac{1}{M}$ the classical action for a straight flux tube describes the heavy quark potential. 

When the distance $R$ between the quark and the antiquark is much larger than the intrinsic width $\frac{1}{M}$ of the classical $Z_N$ flux tube, long distance fluctuations of the axis  must be taken into account and give rise to an effective string theory. To leading order in $\frac{1}{M}$ the action of the effective string theory $S_{eff} (\tilde {x})$ is the classical action for a curved vortex sheet $\tilde{x}$, evaluated in the limit $\frac{1}{M} \rightarrow 0$. This action is equal to the Nambu-Goto action with  a string tension given by the energy per length of an infinite straight flux tube.

\subsection {The Constraint  $\tau=0$ and its Consequences} 
\label{Consequences}

We have introduced a new long-distance parameter, the torque per unit length $\tau$ (Eq (\ref{tau})), linking the string tension to the  chromoelectric field on the axis of an infinite straight flux tube (Eq (\ref{sigmaWprime})).   For large $R$,  the parameter $\tau$ determines the difference between the chromoelectric field  of a quark-antiquark pair  at the positions $|z| = \pm \frac{R}{2}$ of the quarks  and its value at  points $|z| \ll \frac{R}{2} $ in the middle of the flux tube. (Eq (\ref{diffEclass})) 

In this paper we have assumed the value  $\tau=0$  characterizes the QCD flux tube. Under this constraint the chromodynamic fields $\vec{E}$ and $\vec{B}$ on a curved vortex sheet $\tilde{x}$  are determined, to leading order in $\frac{1}{M}$, 
in terms of the string tension (Eq (\ref{newEandBclass})), and the Nambu-Goto action is expressed in terms of these fields  on $\tilde{x}$ (Eq (\ref{Wtauprime})). 

 Imposition of the condition $\tau=0$  on the flux tubes in a particular
$SU(3)$ model  \cite{BBZ:1990} gives a physical picture of these flux tubes in which the behavior of the moment of the pressure (Eq (\ref{pressure})) 
defines a boundary separating a repulsive
interior from an attractive exterior (Fig. \ref{3Ttth}).

 \subsection {Testing the Constraint}
 \label{Test}
 
Testing our conjecture is a problem. (The fit of early lattice simulations of heavy quark potentials and flux tube energy distributions to classical calculations of these quantities discussed in section \ref{specific}  kept $\kappa$ fixed and therefore provides only a crude test.)

Recent lattice simulations \cite{pedro, simulations} of field and energy distributions can be used to test the consistency of the condition $\tau=0$, taking into account string fluctuations in the interpretation of the lattice data.
The results of these simulations can be compared with the
relations (\ref{newEandBclass}) expressing the fields on the vortex sheet
in terms of the string tension, generalizing (\ref{newsigma})
to curved sheets. Comparison with lattice data, of the predicted ratio of 
the field at the center of a flux tube to the string tension, 
provides the most direct test of the
constraint (\ref{balance}). Further lattice data and analysis is necessary
to put strong limits on  $\tau$.
Testing our conjecture is a problem that remains to be solved.



\subsection {Discussion}
\label{Discussion}

According to the correspondence (\ref{newEandBclass}), the world sheet variables $\tilde{x}^\mu$ of effective string theory are associated with  chromodynamic fields of effective magnetic gauge theory on this sheet.  This association, combined with the correspondence of these fields with the underlying fields of Yang Mills theory \cite{Nora}, provides a relation between effective string theory and long distance Yang Mills theory.

The location of the string can thus  be thought of as the axis of a classical flux tube, and the fields associated with the string  regarded as
the  classical chromodynamic fields on that axis.  
We have shown that this is possible  if the flux tube  structure is constrained by the condition $\tau=0$.  Our conjecture is that the  QCD flux tube has the requisite structure.

 We have obtained the constraint  $\tau=0$  on the structure of the QCD flux tube by requiring that no  field energy is created by the separation of a quark-antiquark pair.
In this situation the heavy quark potential, the energy available for doing work when the pair is released, is equal to the energy available when the pair is released in the fixed field created by their separation.   The string tension is then equal to the charge on the quark multiplied by the field on the axis of an infinite flux tube; i. e., the field of the "string" connecting the quark and antiquark.

\vspace{0.3cm}


\section {Acknowledgments}

I would like to thank  P. Bicudo, S. Dubovsky,  and  L. Yaffe for very helpful discussions.

\end{document}